%% file: main.tex
\documentclass[letterpaper,twocolumn,10pt,usenames,dvipsnames]{article}
\usepackage{usenix2019_v3}

\PassOptionsToPackage{usenames,dvipsnames}{xcolor}
\usepackage{listings}
\makeatletter
\let\old@lstKV@SwitchCases\lstKV@SwitchCases
\def\lstKV@SwitchCases#1#2#3{}
\makeatother
\usepackage{lstlinebgrd}
\makeatletter
\let\lstKV@SwitchCases\old@lstKV@SwitchCases

\lst@Key{numbers}{none}{%
    \def\lst@PlaceNumber{\lst@linebgrd}%
    \lstKV@SwitchCases{#1}%
    {none:\\%
     left:\def\lst@PlaceNumber{\llap{\normalfont
                \lst@numberstyle{\thelstnumber}\kern\lst@numbersep}\lst@linebgrd}\\%
     right:\def\lst@PlaceNumber{\rlap{\normalfont
                \kern\linewidth \kern\lst@numbersep
                \lst@numberstyle{\thelstnumber}}\lst@linebgrd}%
    }{\PackageError{Listings}{Numbers #1 unknown}\@ehc}}
\makeatother

\usepackage{hyperref}
\usepackage{breakurl}
\usepackage{changes}
\usepackage[utf8]{inputenx}
\usepackage[T1]{fontenc}
\usepackage{cite}
\usepackage{amsmath,amssymb,amsfonts}
\usepackage{marvosym}
\usepackage{graphicx}
\usepackage{textcomp}
\usepackage{color}
\usepackage{url}
\usepackage{subcaption}
\usepackage{paralist}
\usepackage{wrapfig}
\usepackage{multirow}
\usepackage{listings}
\usepackage{booktabs}
\usepackage{makecell}
\usepackage{boxedminipage}
\usepackage{amsmath}
\usepackage{graphicx}
\usepackage{minibox}
\usepackage{algorithm2e}
\usepackage{algpseudocode}
\usepackage{pifont}
\usepackage{caption}
\usepackage{fancyhdr}
\usepackage{lipsum}
\usepackage{enumitem}
\usepackage{subfiles}
\usepackage[english]{babel}
\usepackage[]{ifsym}
\usepackage[english]{babel}
\usepackage{amsthm}

\usepackage{tikz}
\usepackage{stmaryrd}
\usepackage{mathtools}
\usepackage{cleveref}
\usepackage[framemethod=TikZ]{mdframed} 
\usepackage{MnSymbol}

\crefformat{section}{\S#2#1#3} 
\crefformat{subsection}{\S#2#1#3}
\crefformat{subsubsection}{\S#2#1#3} 

\newcolumntype{H}{>{\setbox0=\hbox\bgroup}c<{\egroup}@{}}

\usepackage{lipsum,afterpage,refcount}

\usepackage[normalem]{ulem}
\usepackage{mdframed}

\newmdtheoremenv{theo}{Theorem}

\usepackage{soul}
\usepackage{tcolorbox}
\tcbuselibrary{theorems}
\tcbset{colback=green!5,colframe=green!35!black,fonttitle=\bfseries}

\usepackage{tabularx} 
\usepackage[nounderscore]{syntax} 

\usepackage{inconsolata} 
\usepackage[italic]{mathastext}

\let\oldcenter\center
\let\oldendcenter\endcenter
\renewenvironment{center}{\setlength\topsep{0pt}\oldcenter}{\oldendcenter}
\usepackage{verbatim}

\begin{document}
\input{macro}


\title{Bridge the Future: High-Performance Networks in Confidential VMs without Trusted I/O devices}
\author{
{\rm Mengyuan Li}\\
MIT\\
lmy@mit.edu
\and
{\rm Shashvat Srivastava}\\
MIT\\
shashvat@alum.mit.edu
\and 
{\rm Mengjia Yan}\\
MIT\\
mengjiay@mit.edu
} 
\maketitle


\begin{abstract}  
Trusted I/O (TIO) is an appealing solution to improve I/O performance for confidential VMs (CVMs), with the potential to eliminate broad sources of I/O overhead.
However, this paper emphasizes that not all types of I/O can derive substantial benefits from TIO, particularly network I/O. Given the obligatory use of encryption protocols for network traffic in CVM's threat model, TIO's approach of I/O encryption over the PCIe bus becomes redundant. Furthermore, TIO solutions need to expand the Trusted Computing Base (TCB) to include TIO devices and are commercially unavailable. 

Motivated by these insights, the goal of this paper is to propose a software solution that helps CVMs immediately benefit from high-performance networks, while confining trust only to the on-chip CVM. We present \sysname, a software solution crafted from a secure and efficient Data Plane Development Kit (DPDK) extension compatible with the latest version of AMD Secure Encrypted Virtualization (SEV), \aka, Secure Nested Paging (SNP). Our design is informed by a thorough analysis of all possible factors that impact SNP VM's network performance. By extensively removing overhead sources, we arrive at a design that approaches the efficiency of an optimal TIO-based configuration. 
Evaluation shows that \sysname has a performance dip less than 6\% relative to the optimal TIO configuration, while only relying on off-the-shelf CPUs.

\end{abstract}
\maketitle

\maketitle
\input{1-intro-v2}
\input{2-background}
\input{3-motivation}
\input{4-overview}

\input{5-implementation}

\input{6-security}
\input{7-evaluation}

\input{8-discussion}
\input{9-related}
\input{10-conclusion}

\bibliographystyle{plain}
{
\bibliography{main}
}
\appendix
\input{0-appendix}
\end{document}

%% file: macro.tex
\newcommand{\bheading}[1]{{\vspace{4pt}\noindent{\textbf{#1}}}}

\newcommand{\xmark}{\ding{55}}%

\newcommand{\tikzcircle}[2][red,fill=red]{\tikz[baseline=-0.5ex]\draw[#1,radius=#2] (0,0) circle ;}%

\newcommand*\emptycirc[1][0.7ex]{\tikz\draw (0,0) circle (#1);} 
\newcommand*\halfcirc[1][0.7ex]{%
  \begin{tikzpicture}
  \draw[fill] (0,0)-- (90:#1) arc (90:270:#1) -- cycle ;
  \draw (0,0) circle (#1);
  \end{tikzpicture}}
\newcommand*\fullcirc[1][0.7ex]{\tikz\fill (0,0) circle (#1);}

\newcommand{\draft}[1]{\textcolor{blue}{#1}}
\newcounter{note}[section]
\renewcommand{\thenote}{\thesection.\arabic{note}}
\newcommand{\yz}[1]{\refstepcounter{note}{\bf\textcolor{red}{$\ll$YZ~\thenote: {\sf #1}$\gg$}}}
\newcommand{\yinqian}{}
\newcommand{\cgx}[1]{\refstepcounter{note}{\bf\textcolor{green}{$\ll$CGX~\thenote: {\sf #1}$\gg$}}}
\newcommand{\guoxing}{\textcolor{green}{$\square$}}
\newcommand{\yan}{}
\newcommand{\mengyuan}[1]{\refstepcounter{note}{\bf\textcolor{olive}{$\ll$mengyuan~\thenote: {\sf #1}$\gg$}}}

\newcommand{\shashvat}[1]{\refstepcounter{note}{\bf\textcolor{blue}{$\ll$shashvat~\thenote: {\sf #1}$\gg$}}}
\newcommand{\my}{\textcolor{olive}{$\square$}}
\newcommand{\mengjia}[1]{\refstepcounter{note}{\small\bf\textcolor{red}{$\ll$Mengjia~\thenote: {\sf #1}$\gg$}}}
\newcommand{\yuheng}[1]{\refstepcounter{note}{\small\bf\textcolor{purple}{$\ll$Yuheng~\thenote: {\sf #1}$\gg$}}}
\newcommand{\jules}[1]{\refstepcounter{note}{\small\bf\textcolor{red}{$\ll$Jules~\thenote: {\sf #1}$\gg$}}}

\newcommand{\algfetch}{%
\begin{algorithm}[H]
\small
$pc \gets $regs.\texttt{get(}$RIP$\texttt{)}\;
\If{TLB.\texttt{search(pc)}== $False$}
{
    // cache TLB\;
    CR$_3 \gets regs.\texttt{get(}$CR$_3\texttt{)}$\;
    \If{$PWC.\texttt{search(}$CR$_3\texttt{)}== False$}
    {
    // cache PWC\;
    $...$\;
    }
    TLB.insert(pc, page\_walk\_result);
}
$p\_addr \gets TLB.\texttt{get(pc)}$\;
\If{$cache.\texttt{search(}p\_addr\texttt{)}== False$}
{
    // cache instruction\;
    $...$\;
}
\end{algorithm}}

\newcommand{\figurewidth}{\columnwidth}
\newcommand{\secref}[1]{\mbox{Section~\ref{#1}}\xspace}
\newcommand{\secrefs}[2]{\mbox{Section~\ref{#1}--\ref{#2}}\xspace}
\newcommand{\figref}[1]{\mbox{Figure~\ref{#1}}}
\newcommand{\algrref}[1]{\mbox{Algorithm~\ref{#1}}}
\newcommand{\tabref}[1]{\mbox{Table~\ref{#1}}}
\newcommand{\appref}[1]{\mbox{Appendix~\ref{#1}}}
\newcommand{\ignore}[1]{}

\newcommand{\etc}{\textit{etc.}\xspace}
\newcommand{\ie}{\textit{i.e.}\xspace}
\newcommand{\eg}{\textit{e.g.}\xspace}
\newcommand{\aka}{\textit{a.k.a.}\xspace}
\newcommand{\etal}{\textit{et al.}\xspace}
\newcommand{\tabincell}[2]{\begin{tabular}{@{}#1@{}}#2\end{tabular}}

\newcommand{\sysname}{\textsc{Folio}\xspace}

\newcommand{\addr}[1]{\ensuremath{P_{#1}}\xspace}
\newcommand{\offset}[1]{\ensuremath{[O_{#1}]}\xspace}
\newcommand{\ts}[1]{\ensuremath{S_{#1}}\xspace}
\newcommand{\gCR}[1]{gCR3\ensuremath{_{#1}}\xspace}
\newcommand{\gbytes}{\ensuremath{\mathrm{GB}}\xspace}
\newcommand{\mbytes}{\ensuremath{\mathrm{MB}}\xspace}
\newcommand{\kbytes}{\ensuremath{\mathrm{KB}}\xspace}
\newcommand{\bytes}{\ensuremath{\mathrm{B}}\xspace}
\newcommand{\hertz}{\ensuremath{\mathrm{Hz}}\xspace}
\newcommand{\ghertz}{\ensuremath{\mathrm{GHz}}\xspace}
\newcommand{\msecs}{\ensuremath{\mathrm{ms}}\xspace}
\newcommand{\usecs}{\ensuremath{\mathrm{\mu{}s}}\xspace}
\newcommand{\nsecs}{\ensuremath{\mathrm{ns}}\xspace}
\newcommand{\secs}{\ensuremath{\mathrm{s}}\xspace}
\newcommand{\gbits}{\ensuremath{\mathrm{Gb}}\xspace}

\newcounter{packednmbr}
\newenvironment{packedenumerate}{
\begin{list}{\thepackednmbr.}{\usecounter{packednmbr}
\setlength{\itemsep}{0pt}
\addtolength{\labelwidth}{4pt}
\setlength{\leftmargin}{12pt}
\setlength{\listparindent}{\parindent}
\setlength{\parsep}{3pt}
\setlength{\topsep}{3pt}}}{\end{list}}

\newenvironment{packeditemize}{
\begin{list}{$\bullet$}{
\setlength{\labelwidth}{0pt}
\setlength{\itemsep}{2pt}
\setlength{\leftmargin}{\labelwidth}
\addtolength{\leftmargin}{\labelsep}
\setlength{\parindent}{0pt}
\setlength{\listparindent}{\parindent}
\setlength{\parsep}{1pt}
\setlength{\topsep}{1pt}}}{\end{list}}

\newenvironment{methoditemize}{
\begin{list}{$\bullet$m}{
\setlength{\labelwidth}{0pt}
\setlength{\itemsep}{2pt}
\setlength{\leftmargin}{\labelwidth}
\addtolength{\leftmargin}{\labelsep}
\setlength{\parindent}{0pt}
\setlength{\listparindent}{\parindent}
\setlength{\parsep}{1pt}
\setlength{\topsep}{1pt}}}{\end{list}}

\lstdefinestyle{mystyle}{
  basicstyle=\ttfamily\small,
  keywordstyle={[1]\color{magenta}},
  keywordstyle={[2]\color{orange}},
  keywordstyle={[3]\color{green}},
  commentstyle=\color{OliveGreen},
  numbers=left,
  numberstyle=\tiny\color{gray},
  numbersep=3pt,
}
\lstdefinelanguage{mylang} {
    morekeywords={[1]module, method},
    morekeywords={[2]Bool, HistoryT, Addr, Data},
    morekeywords={[3]if},
    sensitive=false,
    morecomment=[l]{//},
}
\lstset{style=mystyle}




\newcommand{\hlc}[2][yellow]{{%
    \colorlet{foo}{#1}%
    \sethlcolor{foo}\hl{#2}}%
}

\mdfsetup{
  backgroundcolor=black!10,
  linecolor=black,
  roundcorner=2mm
}

%% file: 1-intro-v2.tex
\section{Introduction}
\label{sec:intro}
Trusted Execution Environment (TEE)~\cite{confidential_consortium:2022} is a promising solution to confidential computing.
A growing trend among these is the support for Confidential Virtual Machines (CVMs or VM-based TEEs), including AMD Secure Encrypted Virtualization (SEV) series~\cite{kaplan:2017:seves,kaplan:2016:sevWpaper,amd:2020:snp}, ARM Confidential Compute Architecture (CCA)~\cite{arm:2021:cca}, and Intel Trust Domain Extensions (TDX)~\cite{intel:2020:tdx}. It is noteworthy that major Cloud Service Providers (CSPs) such as AWS~\cite{aws:2023:sev}, Google Cloud~\cite{google:2020:sev}, and Microsoft Azure~\cite{microsoft:2021:sev}, have officially rolled out commercial services leveraging SEV for CVMs, marking the formal commercialization of VM-based TEE.


\subsection{Motivation: Network I/O with TIO}

\begin{figure}
\centering
\begin{subfigure}[b]{0.302\columnwidth}
\includegraphics[width=\textwidth]{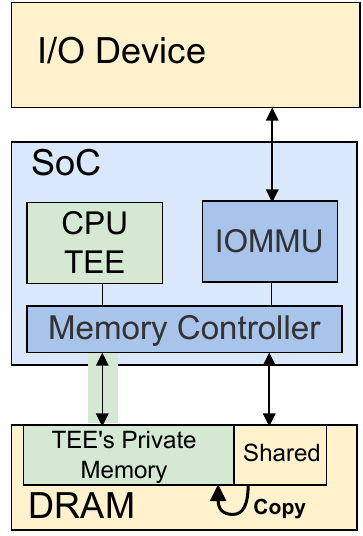}
\caption{\footnotesize Traditional I/O.}
\label{fig:io_overview}
\end{subfigure}
\begin{subfigure}[b]{0.43\columnwidth}
\hspace*{18pt}
\includegraphics[width=\textwidth]{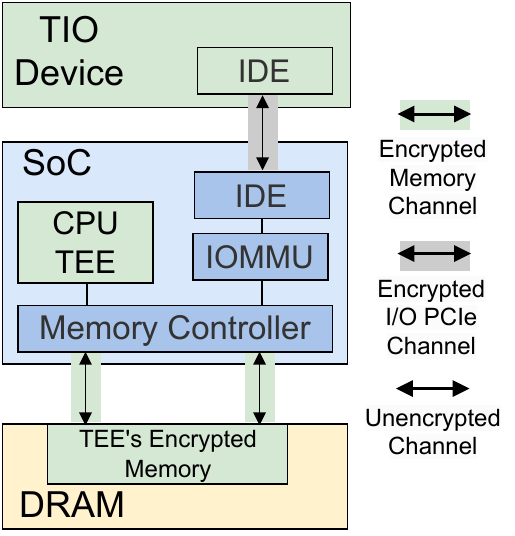}
\caption{\footnotesize Trusted I/O with TIO.}
\label{fig:tio_overview}
\end{subfigure}
\vspace{-5pt}
\caption{Two I/O data flow in confidential VMs. }
\vspace{-5pt}
\label{fig:io_comparison}
\end{figure}

However, CVMs still grapple with suboptimal I/O performance~\cite{akram:2022:sok,akram:2021:performance}.
There are mainly two reasons that count for this: (1) Overhead due to TEE's various security measures that generates broad general side effect, such as additional TLB flushes~\cite{li:2021:tlb} and ownership check~\cite{amd:2020:snp}. 
(2) Overhead due to the fact that I/O devices fall outside the security boundary of CVMs, necessitating additional secure check or software encryption. 
For instance, as shown in \figref{fig:io_overview}, the private memory of CVMs is usually encrypted~\cite{intel:2020:tdx,kaplan:2016:sevWpaper}, to prohibit untrusted devices from accessing private memory through Direct Memory Access (DMA). 
Instead, data is copied to an unencrypted shared memory region to make it accessible to DMA operations. Moreover, software-based encryptions, such as full-disk encryption or encrypted packets, are needed for I/O data.


To solve poor I/O performance, the industry has laid a roadmap for high-performance TEE I/O. Both Intel and AMD recently published whitepapers~\cite{intel:2023:tio,amd:2023:tio} to introduce a new hardware standard, called trusted I/O (TIO), which could be generally applied to any I/O device, including network interface cards (NICs), storage disks, and GPUs.  
Notably, the remote attestation's capability of TIO devices allows the CPU and other devices to function as a cohesive security domain, thereby eliminating most I/O overhead caused the TEE security measures and isolation mechanisms.
Concretely, as shown in \figref{fig:tio_overview}, they now share the same memory encryption key, enabling direct DMA. Additionally, the PCI Integrity and Data Encryption (IDE) protocol also encrypts I/O stream over PCIe bus. All of these operations are further accelerated by hardware, resulting in confidential and high-performance I/O. 

However, network I/O exhibits some fundamental differences compared to other types of I/O, which may limit the extent to which TIO devices can benefit performance. In the case of disk or GPU, the I/O data can be stored, processed, and protected locally within the respective TIO devices. As a result, TIO can gain significant performance improvement by eliminating I/O overhead due to the slow and software-based encryption. In contrast, networks operate differently. \textit{In TEE's threat model, network packets must undergo secure encryption protocols (e.g., TLS or IPsec) before leaving NICs. Consequently, encrypting network packets once, in a protocol-chosen manner, should suffice to secure the data flow between CVMs and NICs.} 
TIO's redundant PCIe encryption step, while not enhancing security, instead introduces unnecessary overhead.
Building upon this insight, this paper raises the following research question:


\vspace{2pt}
\begin{center}
\minibox[rule=1pt,pad=1pt]{
\begin{minipage}[h]{0.9\columnwidth}
{\it \textit{
Can we achieve the same performance as a TIO network solution, but only relying on CPU functionalities, thereby drastically reducing trust assumptions?}}
\end{minipage}
}
\end{center}
\vspace{3pt}

We believe that exploring how to achieve high-speed and secure network in conventional CVMs, where the TCB is strictly enclosed only within CPUs, will remain highly valuable in the coming years, serving as a crucial \textit{\textbf{bridge}} towards the future TIO solution. 
This is especially important at the current stage, considering the pressing demand for high-speed networks, while TIO may not be widely available soon due to the lengthy hardware product cycles. More importantly, directly adopting a TIO solution carries inherent risks by
including peripheral devices within the TCB, which emphasizes the need for a measured approach.  
In this paper, we specifically focus on evaluating the network performance of the latest version of AMD SEV (SEV-SNP), which is the only commercially available CVM at the time of writing. 


\subsection{Comprehensive Analysis of the Sources of Network Performance Overhead in SNP}
To design a software-based solution aimed at minimizing network overhead without TIO devices, a series of exploratory experiments are first carried out to thoroughly investigate the impact of overhead factors that influence network performance in CVMs. 
These factors encompass two main categories: factors due to virtualization or TEE-specific measures.

Our analysis leads to two insightful discoveries.
First, the routing overheads due to virtualization are the dominant factors across all overhead sources, while the TEE-specific overhead plays a marginal role. 
As the TIO device can function in the same way as with non-TEE VMs, these dominant routing overheads can be naturally mitigated by combining TIO devices together with existing optimization techniques, such as Single Root I/O Virtualization (SR-IOV)~\cite{dong:2012:sriov} and the Data Plane Development Kit (DPDK)~\cite{dpdk:2023:dpdk}. 
Therefore, we consider a plausible future configuration that utilizes TIO, SR-IOV, and DPDK together as an optimal setup, ultimately achieving the upper-bound \textbf{\textit{Projected Optimal Network Performance}}.


Another interesting observation is that the overhead introduced by memory encryption may be less than 2\%. As this is the main overhead that cannot be solved by software, such discovery further shows the possibility of \textit{closing} the gap with the projected optimal network configuration. 
By methodically checking off each influencing source, we have found that by carefully dealing with encrypted memory and adding DPDK support to SNP, we can address the majority of factors due to SEV-specific protections. By further minimizing the overhead due to memory encryption, we can close the gap to the utmost degree using software and CPU-only resources. 



\subsection{\sysname: Secure and Performant DPDK Extension for SNP VMs}
We then design and implement \sysname, short for Fast, Optimized, Locked (protected) I/O.
\sysname is a DPDK-based software solution specifically designed for AMD SEV-SNP.
The goal of \sysname is to achieve a performance that matches the optimal network configuration while maintaining security.

To achieve security, \sysname meticulously manages two interfaces: the one between the I/O device and the VM, and the one between DPDK library and DPDK applications. 
This generic approach is shown to guarantee security while maintaining code compatibility.
To achieve efficiency, we introduce a shadow memory design to coordinate secure and fast communication over shared memory, which works together with DPDK optimizations to mitigate most overheads.
Furthermore, \sysname can support network-boosting functionalities, such as packet encryption offloading, using on-core accelerators.
The performance of \sysname is evaluated under different network loads. We also estimate the performance gap between \sysname and projected optimal network performance. 




The contributions of our work are summarized as:
\begin{packeditemize}
\item This paper assesses potential inefficiencies of TIO devices on network, prompting the development of a software solution with a reduced TCB size and comparable performance.

\item This paper investigates the impact of AMD SEV's security measures on network performance.

\item This paper proposes \sysname, a DPDK-based network solution tailored for SNP VMs, which achieves high-speed network performance comparable to the projected optimal TIO solution without compromising existing I/O security.

\item This paper estimates the performance gap between \sysname and the optimal solution. The results indicate that the performance gap between them is within 6\%.

\item This paper extends an existing DPDK evaluation tool to support evaluation of IPsec in confidential VM environments.

\end{packeditemize}



%% file: 2-background.tex
\vspace{-5pt}
\section{Background}
\label{sec:background}
\definecolor{highlightyellow}{RGB}{255,255,204} 
\definecolor{highlightgreen}{RGB}{204,255,204} 

Network performance overhead can come from various factors in TEE systems. 
In this section, we review existing techniques for designing CVMs and survey the potential factors influencing network performance. 
Specifically, we will highlight all factors that affect network performance (\colorbox{highlightyellow}{\textit{Yellow}}represents SEV-specific or TIO-specific factors, and \colorbox{highlightgreen}{\textit{Green}}represents common virtualization factors). 

\vspace{-5pt}

\subsection{Confidential VM}
Confidential VM (CVM), also known as VM-based TEE, is a TEE design used to protect the entire VM. 
Compared to enclave-based TEEs, which are designed to protect an application, CVMs offer better performance and code compatibility. 
At the time of writing, AMD EPYC server is the only off-the-shelf CPU that supports CVMs (AMD SEV). Therefore, this paper will focus on evaluating and analyzing AMD SEV as a representative of CVMs.

\bheading{Memory encryption.} 
Most VM-based TEEs use memory encryption to protect memory and have a memory encryption engine to automatically encrypt data using each VM's unique encryption key.
However, it also introduces two potential impacts on network performance. 
Firstly, all memory accesses now require decryption first (\colorbox{highlightyellow}{\textit{Encrypted memory overhead}}).
Secondly, I/O devices cannot directly access the VM's encrypted private memory for I/O operations, such as Direct Memory Access (DMA).
Instead, a shared region, called bounce buffer, is utilized to facilitate I/O data transmission.
Specifically, the VM kernel needs to actively allocate memory segments from the bounce buffer region and then copy data between the segments and its private memory for each I/O operation, resulting in \colorbox{highlightyellow}{\textit{Bounce buffer overhead}}.

\bheading{Trusted I/O devices.} Trusted I/O (TIO) devices are specifically designed to include devices inside CVM's security boundary and thus address I/O overhead due to security measures from the design of CVMs.
For example, 
the TEE Device Interface Security Protocol (TDISP) defined by TIO standard~\cite{amd:2023:tio,intel:2023:tio} allows the TIO devices to share the memory encryption key used by the VM and directly access the private memory of the VM. 
Moreover, TIO solutions offer protection between the VM and the device by introducing additional hardware-accelerated I/O data link encryption over the PCIe bus (\colorbox{highlightyellow}{\textit{I/O PCIe encryption overhead}}).

\subsection{Evolution of AMD SEV}
AMD SEV has three versions: SEV, SEV-ES, and SEV-SNP. Each subsequent version builds upon the previous one by adding more protective measures and patching known attacks.

\bheading{AMD SEV.} The baseline AMD SEV protects the VM by isolating data within the CPU and using memory encryption to secure data outside the CPU. However, SEV encounters a fatal design flaw: during \texttt{VMEXIT}, the VM's register values are stored in plaintext within the Virtual Machine Control Block (VMCB). This vulnerability directly exposes the register to attackers~\cite{hetzelt:2017:security, werner:2019:severest}, enabling leakage and tampering.

\bheading{AMD ES.} SEV Encrypted State (ES) addresses the exposed register vulnerability by encrypting the register values during \texttt{VMEXIT} (\colorbox{highlightyellow}{\textit{Register encryption overhead}}).
Additionally, the CPU also calculates a measurement for registers and checks for any tampering or replay during subsequent \texttt{VMRUN} operations (\colorbox{highlightyellow}{\textit{\texttt{VMEXIT} check overhead}}). However, as the register values are no longer directly exposed to the hypervisor, for instructions that require hypervisor emulation, such as \texttt{CPUID} and \texttt{RDTSCP}, ES introduces a guest-host communication protocol~\cite{amd:2020:ghcb} to selectively pass register values. A specialized handler called VMM Communication (VC) handler within the VM actively copies the register value that needs to be exposed to the host into a shared area. After emulation, the values are copied back (\colorbox{highlightyellow}{\textit{VC handler overhead}}).
SEV-ES primarily faces two types of attacks. The first attack is related to the unprotected nested page tables used by SEV~\cite{li:2020:crossline,morbitzer:2019:extract_new,morbitzer:2018:severed}.
The second attack is the TLB poisoning attack~\cite{li:2021:tlb}, where the hypervisor can skip some TLB flush. 

\bheading{AMD SNP.} SEV Secure Nested Paging (SNP) addresses the unprotected page table flaw by introducing integrity protection.
Specifically, SNP hardware utilizes a Reverse Map Table (RMP) data structure to store ownership and the corresponding mapping information for each memory page. Now, for each memory access, the hardware compares these metadata to prohibit unauthorized access to VM's private memory (\colorbox{highlightyellow}{\textit{Ownership check overhead}}). 
Note that such protection can also prevent the hypervisor from tampering with memory pages containing encrypted registers. Consequently, SNP VMs no longer face \textit{\textbf{``\texttt{VMEXIT} check overhead"}}.
For the TLB poisoning attack, SNP checks whether a TLB flush should be enforced during each \texttt{VMRUN} (\colorbox{highlightyellow}{\textit{TLB check overhead}}).


\subsection{Existing Network Techniques }
There are three common network techniques used in VM.

\bheading{VirtIO.} VirtIO~\cite{russell:2008:virtio} is the most common and default networking configuration in Kernel-based Virtual Machine (KVM), including AMD SEV (where KVM and Quick Emulator (QEMU) are used). 
When VirtIO is enabled, the hypervisor emulates virtual I/O devices for the VM. The VM interacts with these emulated devices to process its I/O traffic. 
In a typical VirtIO setup with KVM, there are three potential factors that can slow down network performance.

Firstly, applications usually still utilize the system call interface to send packets, which introduces overhead due to user-kernel context switches and the VM kernel's handling of packet processing (\colorbox{highlightgreen}{\textit{Routing within the VM overhead}}).
Secondly, the emulated network devices introduce overhead on the host side (\colorbox{highlightgreen}{\textit{Routing within the host overhead}}).
Thirdly, all I/O interrupts are emulated in KVM (\colorbox{highlightgreen}{\textit{Emulated I/O interrupt overhead}}). 
Physical I/O interrupts are not delivered directly to the VM, but are injected into the corresponding vCPU in an emulated way. 

\bheading{Single Root I/O Virtualization (SR-IOV).} SR-IOV is an extension of the PCIe standard and can help to solve the overhead due to routing within the host side. SR-IOV allows I/O devices, such as a smart NIC, to share physical resources among a Physical Function (PF) and several Virtual Functions (VFs) and binds each function directly to a VM. For example, in an Intel 82599 10GB NIC, up to 63 VFs can be supported per network port. 
The hypervisor can then bind one or more VFs to the VM directly using PCI passthrough. This helps bypass the overhead introduced by device emulation layers like VirtIO, enabling the VM to mitigate overhead due to \textit{\textbf{``Routing within the host overhead"}}. 

\bheading{Data Plane Development Kit (DPDK).} DPDK is a software framework designed to boost network performance by mitigating overhead due to \textit{\textbf{``Routing within VM"}} and \textit{\textbf{``Emulated I/O interrupts"}}. 
DPDK provides a series of userspace APIs through the construction of an Environment Abstraction Layer (EAL) to assist DPDK applications in fast network processing. 
DPDK takes advantage of a range of techniques hidden within the EAL to accelerate network, including: (1) Userspace implementation, where DPDK applications can run without switching to kernel space, avoiding the overhead of internal routing. (2) Zero copy, where network packets are directly passed to the memory of DPDK applications, eliminating the need to copy between user space and kernel space. (3) Poll Mode Driver (PMD), which allows DPDK applications to send and receive packets through polling, without relying on interrupts. (4) Hardware-based optimizations, such as cache alignment and the utilization of huge pages. 


%% file: 3-motivation.tex
\section{Identifying Network Bottlenecks in CVMs}
\label{sec:motivation}

\begin{table*}[t]
\setlength{\tabcolsep}{3pt}
\centering
\scriptsize
\scalebox{1}{

    \begin{tabular}{cl|c|c|c|c|c|c|c|c|c|c}
        \hline\hline
        &&\multicolumn{3}{c|}{\hspace{11pt}\textbf{non-TEE VM}}&\multicolumn{4}{c|}{\textbf{SEV/ ES/ SNP Configurations}}&\multicolumn{3}{c}{\textbf{Proposed Configurations}}\\ 
        \cline{3-12} 
        \\[-7pt]
        
        &\textbf{\tabincell{c}{Overhead Description}}	& 	\tabincell{c}{non-TEE VM}	&
        \tabincell{c}{VM with \\ SRIOV}	&
        \tabincell{c}{VM with \\ DPDK}	&
        SEV VM	&  \tabincell{c}{SEV with \\ SRIOV} &
	\tabincell{c}{ES VM \\with SRIOV}		& \tabincell{c}{SNP VM \\with SRIOV}& \tabincell{c}{SNP VM \\with TIO}& \tabincell{c}{SNP VM \color{LimeGreen}$\filledstar$ \\(TIO/ DPDK)}& \tabincell{c}{SNP VM \\(\sysname)}\\\hline
                \\[-7.5pt]
                
        \multicolumn{1}{c}{\multirow{4}{*}{\rotatebox{90}{\texttt{\textbf{Common}}}}}	& \multicolumn{1}{|l|}{\texttt{Routing within the VM}}	&	\color{Maroon}\textbf{Y}	&	\color{Maroon}\textbf{Y}	&	\color{LimeGreen}\textbf{N}	&	\color{Maroon}\textbf{Y} & \color{Maroon}\textbf{Y}	&	\color{Maroon}\textbf{Y}		&	\color{Maroon}\textbf{Y}	&	\color{LimeGreen}\textbf{N}	&	\color{LimeGreen}\textbf{N}	&	\color{LimeGreen}\textbf{N} \\
        
        &\multicolumn{1}{|l|}{\texttt{Routing within the host}}	&	\color{Maroon}\textbf{Y}	&	\color{LimeGreen}\textbf{N}	&	\color{LimeGreen}\textbf{N}	&	\color{Maroon}\textbf{Y} & \color{LimeGreen}\textbf{N}	&	\color{LimeGreen}\textbf{N}	&	\color{LimeGreen}\textbf{N}	&	\color{LimeGreen}\textbf{N}	&	\color{LimeGreen}\textbf{N}	&	\color{LimeGreen}\textbf{N}	\\
        
        &\multicolumn{1}{|l|}{\texttt{Emulated I/O interrupt}}	&	\color{Maroon}\textbf{Y}	&	\color{Maroon}\textbf{Y}	&	\color{LimeGreen}\textbf{N}	&	\color{Maroon}\textbf{Y} & \color{Maroon}\textbf{Y}	&	\color{Maroon}\textbf{Y}		&	\color{Maroon}\textbf{Y}	&	\color{Maroon}\textbf{Y}	&	\color{LimeGreen}\textbf{N} &	\color{LimeGreen}\textbf{N}	\\
        
        &\multicolumn{1}{|l|}{\texttt{Other factors}}	&	\color{Maroon}\textbf{Y}	&	\color{Maroon}\textbf{Y}	&	\color{Maroon}\textbf{Y}	&	\color{Maroon}\textbf{Y} & \color{Maroon}\textbf{Y}	&	\color{Maroon}\textbf{Y}		&	\color{Maroon}\textbf{Y}	&	\color{Maroon}\textbf{Y}	&	\color{Maroon}\textbf{Y}    &	\color{Maroon}\textbf{Y}	\\
        \hline 
        \\[-7pt]
        \multicolumn{1}{c}{\multirow{9}{*}{\rotatebox{90}{\texttt{\textbf{SEV-only}}}}}	&\multicolumn{1}{|l|}{\texttt{Encrypted memory overhead}}	&	\color{Maroon}\textbf{Y}	&	\color{Maroon}\textbf{Y}	&	\color{Maroon}\textbf{Y}	&	\color{Maroon}\textbf{Y} &	\color{Maroon}\textbf{Y} &	\color{Maroon}\textbf{Y}		&	\color{Maroon}\textbf{Y}	&	\color{Maroon}\textbf{Y}	&	\color{Maroon}\textbf{Y}   &	\color{Maroon}\textbf{Y}	\\
        
        &\multicolumn{1}{|l|}{\texttt{Register encryption }}	&	\color{Maroon}\textbf{Y}	&	\color{Maroon}\textbf{Y}	&	\color{LimeGreen}\textbf{N+}	&	\color{Maroon}\textbf{Y} &	\color{Maroon}\textbf{Y} &	\color{Maroon}\textbf{Y}	&	\color{Maroon}\textbf{Y}	&	\color{Maroon}\textbf{Y}	&	\color{LimeGreen}\textbf{N${^+}$}   &	\color{LimeGreen}\textbf{N${^+}$}	\\
        
        &\multicolumn{1}{|l|}{\texttt{Bounce buffer allocation}}	&	\color{LimeGreen}\textbf{N}	&	\color{LimeGreen}\textbf{N}	&	\color{LimeGreen}\textbf{N}	&	\color{Maroon}\textbf{Y} &	\color{Maroon}\textbf{Y} &	\color{Maroon}\textbf{Y}	&	\color{Maroon}\textbf{Y}	&	\color{LimeGreen}\textbf{N}	&	\color{LimeGreen}\textbf{N} &	\color{LimeGreen}\textbf{N}	\\

        &\multicolumn{1}{|l|}{\texttt{Bounce buffer copy}}	&	\color{LimeGreen}\textbf{N}	&	\color{LimeGreen}\textbf{N}	&	\color{LimeGreen}\textbf{N}	&	\color{Maroon}\textbf{Y} &	\color{Maroon}\textbf{Y} &	\color{Maroon}\textbf{Y}	&	\color{Maroon}\textbf{Y}	&	\color{LimeGreen}\textbf{N}	&	\color{LimeGreen}\textbf{N} &	\color{Maroon}\textbf{Y}	\\
        
        &\multicolumn{1}{|l|}{\texttt{VMEXIT check overhead}}	&	\color{LimeGreen}\textbf{N}	&	\color{LimeGreen}\textbf{N}	&	\color{LimeGreen}\textbf{N}	&	\color{LimeGreen}\textbf{N}	& \color{LimeGreen}\textbf{N} &	\color{Maroon}\textbf{Y}		&	\color{LimeGreen}\textbf{N}	&	\color{LimeGreen}\textbf{N}	&	\color{LimeGreen}\textbf{N}	&	\color{LimeGreen}\textbf{N}	\\
        
        &\multicolumn{1}{|l|}{\texttt{VC handler overhead}}	&	\color{LimeGreen}\textbf{N}	&	\color{LimeGreen}\textbf{N}	&	\color{LimeGreen}\textbf{N}	&	\color{LimeGreen}\textbf{N}	& \color{LimeGreen}\textbf{N} &	\color{Maroon}\textbf{Y}		&	\color{Maroon}\textbf{Y}	&	\color{Maroon}\textbf{Y}	&	\color{LimeGreen}\textbf{N}$^*$	&	\color{LimeGreen}\textbf{N}$^*$	\\
        
        &\multicolumn{1}{|l|}{\texttt{Ownership check overhead}}	&	\color{LimeGreen}\textbf{N${^-}$}	&	\color{LimeGreen}\textbf{N${^-}$}	&	\color{LimeGreen}\textbf{N${^-}$}	&	\color{LimeGreen}\textbf{N${^-}$}	& \color{LimeGreen}\textbf{N${^-}$} &	\color{LimeGreen}\textbf{N${^-}$}		&	\color{Maroon}\textbf{Y}	&	\color{Maroon}\textbf{Y}	&	\color{Maroon}\textbf{Y}    &	\color{Maroon}\textbf{Y}	\\

        &\multicolumn{1}{|l|}{\texttt{TLB check overhead}}	&\color{LimeGreen}\textbf{N}	&	\color{LimeGreen}\textbf{N}	&	\color{LimeGreen}\textbf{N}	&\color{LimeGreen}\textbf{N}	& \color{LimeGreen}\textbf{N} &\color{LimeGreen}\textbf{N}	&\color{Maroon}\textbf{Y}	&\color{Maroon}\textbf{Y}	&	\color{LimeGreen}\textbf{N${^+}$}   &	\color{LimeGreen}\textbf{N${^+}$}	\\

        &\multicolumn{1}{|l|}{\texttt{I/O PCIe encryption overhead}}	&	\color{LimeGreen}\textbf{N}	&	\color{LimeGreen}\textbf{N}	&	\color{LimeGreen}\textbf{N}	&	\color{LimeGreen}\textbf{N} &	\color{LimeGreen}\textbf{N} &	\color{LimeGreen}\textbf{N}	&	\color{LimeGreen}\textbf{N}	&	\color{Maroon}\textbf{Y}	&	\color{Maroon}\textbf{Y} &	\color{LimeGreen}\textbf{N}	\\

	\hline\hline \\[-7pt] \multicolumn{11}{l}{\textbf{\hspace{9pt} Standardized Latency (5000 pps)}} \\\hline	

         &Mean latency   & >50x  & 1.00 (75.1 \textmu s)  & 0.37x & >50x & 1.01x& 1.17x & 1.16x & N/A & N/A & N/A \\
         &Median latency  & >50x & 1.00 (75.3 \textmu s)  & 0.36x & >50x & 1.02x & 1.18x & 1.13x &  N/A & N/A & N/A\\
         &p95 tail latency & >50x &1.00 (83.8 \textmu s)   & 0.33x & >50x & 1.00x & 1.20x & 1.18x  & N/A & N/A & N/A \\
         &p99 tail latency & >50x & 1.00 (123.6 \textmu s) & 0.31x & >50x & 0.78x& 0.94x& 2.19x & N/A & N/A & N/A \\
         
        \hline\hline 
    \end{tabular}
}
\caption{\small Factors affecting network under different VM configurations and the standardized round trip latency. \textcolor{Maroon}{\textbf{Y}} implies that there is this overhead and \textcolor{LimeGreen}{\textbf{N}} implies not. With DPDK means the VM is using DPDK to operate a SR-IOV device. With TIO means the VM is using TIO devices with SR-IOV. $\color{LimeGreen}\filledstar$ means the projected optimal solution. Other factors refer to the impact of physical devices and software settings, which are all configured the same.  - indicates that the ownership check only happens for write access. * indicates that VC handler overheads due to the scheduler or interrupts are minimized. + indicates that the number of \texttt{VMEXIT} and thus such overhead is minimized because of polling mode. For standardized latency, the performance of ``non-TEE VM with SR-IOV" is chosen as the standard with their value represented in parentheses. $x$ represents the ratio of latency under different configurations compared to the standard. N/A means this configuration is not supported yet.}
\label{tab:factors}
\end{table*}

We first conducted an in-depth 
analysis to understand the relative importance of each forementioned overhead factor, which plays a crucial role in guiding us to propose a software solution. Our measurement results indicate that it is possible to use software-only methods to mitigate all major factors.





\subsection{Overview}

\bheading{Evaluation Methodology.} 
\label{sec:motivation:method}
In different VM configurations (such as SEV VM or non-TEE VM), there are distinct factors that affect the network performance. By comparing the performance of two different VM configurations with or without a specific factor, we can estimate the impact of that factor.
\tabref{tab:factors} summarizes influencing factors in each VM configuration, including three proposed configurations that are not supported or available.
Under different VM configurations, we run a simple UDP echo server within the VM and run a client within the local network to measure and compare their average round-trip latency and tail latency.

Using this comparison methodology has two advantages:
(1) It can help estimate the impact of factors that cannot be directly measured.
For example, the ownership check may occur on every memory page access, making it impractical to inject code to measure it.
(2) Eliminate inaccurate results caused by frequently reading timestamps within the VM. Frequently reading timestamps inside VM can significantly impact performance itself, leading to inaccurate results, especially in SEV configurations. This is because inside VM, instructions such as \texttt{RDTSCP} are emulated by default (\eg, trigger \texttt{VMEXIT} and the hypervisor emulates the result), which slows down the application and falsely \textbf{magnifies} the time spent on the measured duration. In settings like SEV-ES and SNP, more complex routing occurs, involving the application switching to the VC handler first, then to the hypervisor, back to the VC handler, and finally back to the application. In a network connection with 5000 packets per second, such intricate routing can introduce significant noise when assessing the dominant factors impacting network performance.

\bheading{SEV's Impacts on Non-TEE VMs.} 
It is worth noting that enabling SEV feature in BIOS also has impacts on the performance of non-TEE VMs.
Specifically, the impacts are reflected in two aspects: (1) Encrypted memory overhead. When enabling SEV features, the Secure Memory Encryption (SME)~\cite{kaplan:2016:sevWpaper} feature is enabled by default. With SME, all memory pages, except CVMs' private pages, are encrypted using the host memory encryption key.
Thus, non-TEE VMs also face encrypted memory overhead and register encryption overhead.
(2) Ownership check. When SNP is enabled, 
partial ownership check is also performed when the hypervisor or non-TEE VMs have memory \textit{write} access, which 
can introduce mild overhead~\cite{li:2022:systematic}. 



\subsection{Experimental Setup}
\label{sec:motivation:setup}
Our experimental setup consists of a SNP-supported workstation and a desktop client. The SNP-supported workstation has an AMD EPYC 7313 16-Core Processor, 64GB DRAM, 1TB disk, an Intel I350 Gigabit NIC for internet connection, and an Intel 82599ES-based 10Gb NIC (Silicom PE210G2SPI9) for supporting DPDK and SR-IOV. The desktop client has an Intel i5-12400F Processor, 32GB DRAM, 500GB disk, an Intel I219-V NIC for sharing internet with the workstation, and another Intel 82599ES-based 10Gb NIC for supporting DPDK and SR-IOV. The two Intel 82599ES NICs are connected by a MokerLink 8 Port 10Gbps switch. The original host kernel (\texttt{sev-snp-iommu-avic\_5.19-rc6\_v4} branch), QEMU (\texttt{snp-v3} branch), and OVMF (\texttt{master} branch) were directly obtained from the \texttt{sev-snp-devel} repository~\cite{amd:2023:github} (Commit: \texttt{fbd1d07628f8a2f0e29e9a1d09b1ac6fdcf69475}). The desktop client simply runs an unmodified Ubuntu 22.04.1 LTS with a kernel of \texttt{5.19.0-38-generic}.

The VMs were configured as 4-virtual CPUs (vCPUs), 8-GB memory, 30GB disk storage. The guest kernel used to support the SNP feature is forked from the same repository as the host kernel. For ``non-TEE VM" and ``SEV VM" setup, the VMs use the default virtio-net-pci device suggested by AMD's official script used for launching SEV VM~\cite{amd:2023:github}. For all SR-IOV and DPDK setups, one NIC's virtual function (VF) is directly assigned to the VM via QEMU PCI pass-through configuration.   
The network server applications running inside the VMs are two simple UDP echo servers. One of them uses normal socket APIs, and the other is only used in ``VM with DPDK" setup and is configured using DPDK APIs. On the client side, we reuse a client application from an existing project that focuses on measuring tail latency~\cite{kaffes:2019:shinjuku}. We configure the packet rate to be 5000 packets per second (pps). 


\subsection{Results and Takeaways} 
By comparing performance for various configurations, we can derive some intriguing observations.

\begin{center}
\begin{minipage}[t]{1\columnwidth}
\vspace{-8pt}
\begin{mdframed}
\textbf{Observation-1:}
Overhead due to routing within the host and routing within the VM are the dominant factors that impact performance.
\end{mdframed}
\end{minipage}
\end{center}
As shown in \tabref{tab:factors}, when focusing on the first group (non-TEE VMs) and comparing whether SR-IOV is enabled, we observed a difference in latency of over 50 times. This indicates that emulated I/O devices and routing within the host side are crucial factors affecting network performance. Similarly, when comparing ``non-TEE VM with SR-IOV" and ``non-TEE VM with DPDK", we can observe that routing within the VM and emulated I/O interrupt may potentially lead to 2-3 times network performance latency.

\begin{center}
\begin{minipage}[t]{1\columnwidth}
\vspace{-8pt}
\begin{mdframed}
\textbf{Observation-2:}
Overhead due to ownership check and TLB check is smaller than that of \texttt{VMEXIT} check. The VC handler maybe the key factor that impact performance for newly SEV-ES and SEV-SNP VMs.
\end{mdframed}
\end{minipage}
\end{center}
The second group is SEV series VMs. When comparing ``ES with SR-IOV" and ``SNP with SR-IOV", we found that despite SNP introducing ownership checks and TLB checks, SNP's performance is better than ES. This could be attributed to the significant performance impact caused by the \texttt{VMEXIT} check in ES. Meanwhile, this also implies that the overhead of ownership and TLB checks is not as substantial.
When comparing ``ES/SNP VMs with SR-IOV" and ``SEV VM with SR-IOV", we noticed that their performance differs by more than 10\%. This suggests that the VC handler might be the primary reason for the performance decline in ES/SNP configurations. Moreover, most of these overheads occur during \texttt{VMEXIT}, and they can be mitigated through the use of DPDK, which effectively reduces the number of context switches.

\begin{center}
\begin{minipage}[t]{1\columnwidth}
\vspace{-8pt}
\begin{mdframed}
\textbf{Observation-3:}
The impact of bounce buffer overhead may be smaller than we previously thought.
\end{mdframed}
\end{minipage}
\end{center}
When comparing these two groups and looking at ``non-TEE VM with SR-IOV" and ``SEV with SR-IOV", their performance difference is only a remarkable 1-2\%. The only distinction between them is the presence of bounce buffer overhead. This strongly suggests that the impact of bounce buffer itself on network performance is likely smaller than what we initially anticipated.

\subsection{Motivation} 


Inspired by \textbf{Observation-3}, we realized despite the fact that the bounce buffer overhead cannot be fully bypassed without TIO devices; such overhead may not be the decisive limiting factor for CVMs, especially if we can further minimize the overhead due to the bounce buffer allocation.

This realization has motivated us to propose a software-based network prototype solution called \sysname, which naturally addresses most SEV-specific overhead by adding secure and efficient DPDK support in SNP, while further reducing the overhead caused by the bounce buffer.
When comparing "SNP VM with \sysname" with the projected optimal solution (``SNP VM with TIO/DPDK"), the only difference lies in one bounce buffer copy overhead. By eliminating all known influencing factors to the greatest extent, we foresee the potential of achieving comparable performance between them.

\bheading{Challenges.} Designing \sysname in SNP poses challenges in terms of \textit{security}, \textit{efficiency}, and \textit{functionality}:
\begin{packeditemize}
\item \textit{Security.} Ensuring security in managing the I/O security boundary (shared memory and private memory) is challenging, especially in DPDK where assistance from the VM kernel is unavailable. It is vital to prevent the VM from leaking its secrets when taking advantage of DPDK, as well as to guarantee the security of existing DPDK applications when running them directly in SNP VMs.
\item \textit{Efficiency.} A significant challenge lies in how to maintain DPDK's performance while efficiently dealing with shared memory regions and private memory regions.
\item \textit{Functionality.} TIO may offer additional crypto offload functionality. Finding ways to implement similar functionality using only the CPU requires careful adjustments.
\end{packeditemize}


%% file: 4-overview.tex
\vspace{-5pt}
\section{\sysname Overview}
\label{sec:overview}
\vspace{-5pt}
In this section, we provide an overview of \sysname, a secure and efficient network solution based on a DPDK extension. 

\subsection{Threat Model}
\sysname operates under the same threat model as the conventional CVM utilizing untrusted I/O devices~\cite{kaplan:2016:sevWpaper,intel:2020:tdx}. In this model, the trusted computing base (TCB) only consists of software running within SNP-protected VMs and hardware components within the SoC. It is also assumed that all software components running inside the VMs are benign and programmed securely. The potential adversary in this threat model typically possesses privileged software control over the entire software stack and has direct physical access to the server machine.
The SNP VMs need to utilize shared memory for interaction with the untrusted hypervisor and I/O devices. Consequently, all data within this shared region, including I/O packets and other metadata, exist in an unprotected state, allowing a malicious hypervisor to eavesdrop, tamper with, or block such data.
Moreover, similar to the threat model of TEE, Denial-of-Service (DoS) attacks are not considered as they can be easily detected by the TEE instance owner, thereby violating the operational model of cloud service providers.


\subsection{Design Goals}
The motivation of \sysname is to achieve secure network transmission while also achieving performance and functionalities that are comparable to the projected optimal solution utilizing TIO. 
Here, we summarize our goals:

\begin{enumerate}[label=\textbf{G\arabic*.},fullwidth,itemindent=0pt,listparindent=\parindent,itemsep=0ex,partopsep=0pt,parsep=0ex]
\item \textbf{End-to-end Security.}\label{goal_security} Achieving the same level of network security as the existing CVMs using VirtIO is the primary goal of \sysname, where VirtIO works together with software-based encryption (\eg, TLS or IPsec) to safeguard network. 
Note that \sysname's security level cannot be directly compared to future TIO solutions, as discussed in \secref{sec:discussion:security}. 

\item \textbf{Comparable Network Performance.}\label{goal_performance} 
Achieve high-performance networking that is on par with the projected optimal network solution (TIO with DPDK) is another goal.


\item \textbf{On-core acceleration of offloading tasks.}\label{goal_offload} 
Although TIO devices are not available, we can anticipate that in situations where the device is deemed trustworthy, CVMs may benefit from offloading network tasks (e.g., packet decryption) to the trusted NIC, thereby enhancing overall performance. Consequently, \sysname needs to support these network performance-enhancing features.

\item \textbf{Code Compatibility.}\label{goal_code} 
Code compatibility with existing DPDK applications provides two potential benefits. Firstly, the existing DPDK source code can be compiled and executed directly within SNP VMs. Secondly, 
\sysname offers a future-proofing for developers aiming to deploy network applications inside CVMs. With the assistance of \sysname, they can validate their implementations and ensure robust performance. Crucially, the effort invested in this process isn't short-lived. When TIO devices come into the picture and gain support, their implementations using \sysname support can direct operate effectively with TIO since they adhere to the same DPDK Application Binary Interface (ABI). This ensures the longevity and sustainability of the code, making the initial investment in testing and development worthwhile.

\end{enumerate}

\subsection{\sysname Design Overview}


To achieve \ref{goal_security} \textbf{End-to-end Security}, \sysname focuses on two essential aspects to ensure security: \textit{Constrained VM-I/O interaction interfaces} and \textit{Constrained DPDK-App interaction interfaces}. These aspects ensure security between the VM and untrusted I/O devices (hypervisor side), as well as security between the DPDK library and DPDK applications.
As shown in \figref{fig:overview}, the interaction between VMs and I/O devices is managed by a combination of a Shared Region Management kernel module and a modified DPDK library. This setup rigorously manages the shared memory region, ensuring that only necessary data is exposed and tightly controlling all network packets and metadata that could be exposed to the untrusted host side. 
Regarding the interaction between the DPDK library and the DPDK applications, we employ a \textit{shadow network buffer pool} design. This design ensures that only data structures within the private memory region can be accessed by DPDK applications. 


To achieve \ref{goal_performance} \textbf{Comparable Network Performance},  
\sysname tries to minimize overhead due to different factors to the greatest extent. This includes using modified DPDK to mitigate most factors associated with SEV-specific protection and using pre-allocated memory segments to mitigate bounce buffer allocation overhead.
These measures enable \sysname to achieve comparable performance with non-TEE VMs.

For \ref{goal_offload} \textbf{On-core acceleration of offloading tasks}, 
\sysname focuses on ensuring end-to-end secure network communication with a specific emphasis on crypto-operation offload support. \sysname leverages AES-NI instruction to accelerate crypto operations, supporting both the look-aside offload mode and a CPU-enabled emulated inline mode (\secref{sec:implementation:ipsec}). To evaluate the performance of network offloading, we also developed an IPsec testing tool
to conduct a comprehensive evaluation in CVMs. 
For \ref{goal_code} \textbf{Code Compatibility}, \sysname maintains the original DPDK ABI, allowing the existing DPDK application to be compiled and executed directly within SNP VMs.



%% file: 5-implementation.tex
\section{\sysname Design Details}
\label{sec:implementation}
This section focuses on the design details \sysname is implemented over DPDK to achieve secure and efficient network.


\subsection{{\fontsize{11pt}{1}\selectfont Constrained VM-I/O Interaction Interface}}



To restrict the interaction between the VM and the I/O device, we employ a \textit{shared region management} module and follow a \textit{limited exposed metadata} principle to ensure only necessary and secure data are exposed.

\bheading{Shared region management.} \sysname introduces a shared region management kernel module to strictly control the contents that need to be placed in shared memory. Specifically, any DPDK-related content that needs to be placed in shared memory must be explicitly notified to this kernel module during DPDK initialization. This module then interacts with the hypervisor to synchronize the corresponding physical addresses of shared regions. When shutting down DPDK applications, the kernel module zeros out all contents placed in these shared regions before recycling memory pages.

\begin{figure}[t]
\centering
\includegraphics[width=0.85\columnwidth]{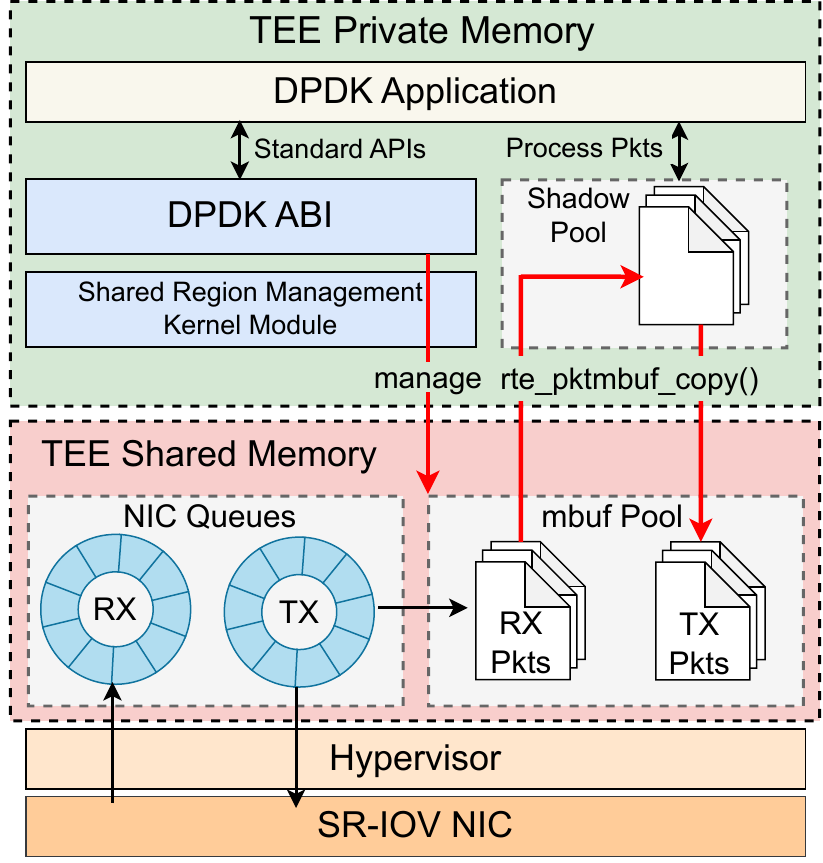}
\caption{\sysname overview. All components within the TEE private memory are trusted. Meanwhile, the hypervisor, NIC, and all data within the shared memory are untrusted. The DPDK library and the shared region management module strictly control VM's interaction with shared memory.}
\label{fig:overview}
\end{figure}

\bheading{Limited exposed metadata.} 
\sysname explicitly and manually examines all data structures exposed to the host.
It turns out that only two types of data region need to be shared. The first is the receive/transmit (RX/TX) descriptor rings, which are also shared in the default VirtIO to facilitate communication with I/O devices. These rings contain metadata information about network packets, such as memory locations or status of network packets. A secure analysis of all exposed metadata is included in \appref{appendix:metadata}. 
The second type of region is the DPDK-specific network buffers (\texttt{rte\_mbuf}), which hold the actual incoming and outgoing network packets and were originally designed to be directly accessible by DPDK applications. We use the shadow packet buffer pool design (\secref{sec:design:app}) to protect this region.
In the manual examination procedure, we conduct a thorough comparison between the data exposed by the default VirtIO and \sysname to ensure that there is no inadvertent information leakage. The most exposed data mainly comprises meta-data associated with packets, which is handled either in an on-core write-only manner or protected by software-based encryption. 
Additionally, we pay extra attention to data structures resembling pointers. For pointers that do not necessarily need to be exposed, such as some DPDK-specific pointer design, we keep them in private memory. For pointers that must be exposed, like addresses in the descriptor rings, we check whether the pointed-to addresses are within the shared region when performing operations.

\subsection{{\fontsize{11pt}{1}\selectfont Constrained DPDK-APP Interaction Interface}}
\label{sec:design:app}
To ensure the security of DPDK applications, \sysname follows a copy-before-processing principle and proposes a shadow packet buffer pool design to ensure that applications only operate on private memory. Additionally, \sysname also conceals shared memory from DPDK applications by hiding vulnerable information (\eg, virtual address of the shared region or exposed data) inside the Environment Abstraction Layer (EAL), and includes special fault handling mechanisms to provide additional protection.

\bheading{Copy before processing.} \sysname strictly adheres to the principle of handling network packets as outlined in VirtIO, \ie, copying the network packet and the associated network buffer data structures from the shared region to the private region before any software begins processing it, and vice versa. Once the network packets are placed in private memory, SNP's memory protection mechanism explicitly ensures the confidentiality and integrity of the packets. This prevents the hypervisor from modifying the packet content or associated data structures of the network packets while an application is using them.
Note that we want to emphasize that even encrypted network packets should be copied to the private region before network applications begin touching them. Some existing applications may attempt to decrypt the packets directly in their original memory addresses or read unencrypted information, such as header information, during packet processing. These actions can enable the attackers to manipulate the VM's intended control flow or provide incorrect data.


\bheading{Shadow packet buffer pool.} 
\sysname introduces a shadow packet buffer pool design to provide a secure interface for DPDK applications to process network packets.  
Specifically, packet buffer pools are the memory pools used by DPDK applications to handle network packets. These pools are created during DPDK initialization, and each pool consists of a fixed number of data structure objects known as \texttt{rte\_mbuf}. These \texttt{rte\_mbuf} objects are responsible for storing network packets and are accompanied by a set of preceding metadata. The metadata includes information such as the message type, length, a region to store some application-specific private data, a pointer to the address of the raw network packet and, if necessary, a pointer to the next \texttt{rte\_mbuf} object in case a single object is insufficient to hold a large network packet.

\begin{figure}[t]
\centering
\includegraphics[width=0.9\columnwidth]{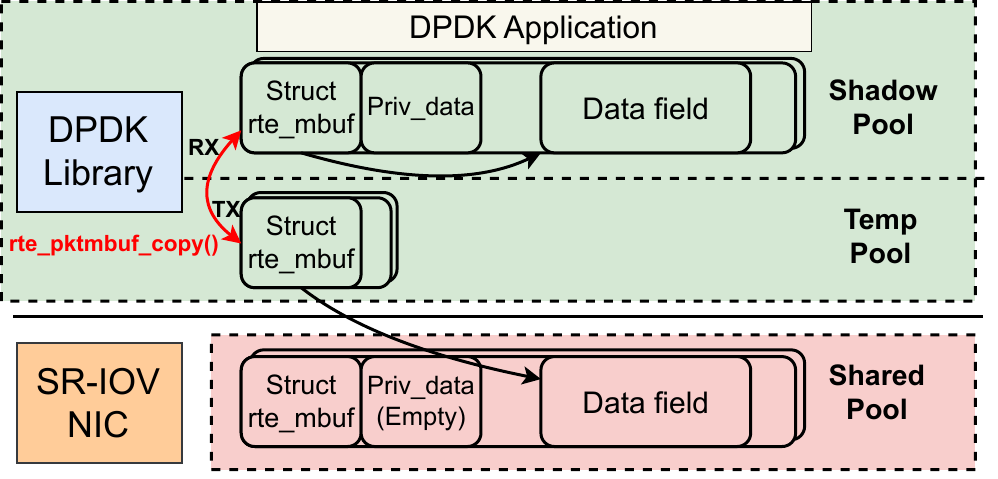}
\caption{Shadow packet buffer pool design.}
\label{fig:pools}
\vspace{-15pt}
\end{figure}

The shadow packet buffer pool design enables the guest VM and DPDK applications to retain critical metadata internally while maintaining the ability to efficiently communicate with I/O devices. 
As shown in \figref{fig:pools}, \sysname creates three packet buffer pools during DPDK initialization, which we call the \textit{shared} pool, the \textit{shadow} pool, and the \textit{temporary} pool. Among these three packet buffer pools, the \textit{shared pool} is the only memory pool allocated in shared memory. It serves as the memory pool used by the untrusted NIC for reading/writing TX/RX packets and is strictly restricted from direct use by the SNP VM. 
The \textit{temporary} memory pool is used by the device driver inside the SNP VM to communicate with the device, where the network packet fields of \textit{temporary} pool point to shared memory while other data structures or metadata are stored safely in private memory. This arrangement allows the device to read and write packets from these memory buffers while keeping other data such as the application's private data and pointers inaccessible to the untrusted devices. 
The \textit{shadow} pool is the memory pool that is fully protected by the private memory and used by the application. 
As a result, DPDK applications can maintain their existing code behaviors without having to worry about source-code modification or any data leakage, including performing in-place packet processing directly in the data field. The memory copy operation between the \textit{shadow} memory pool and the \textit{temporary} memory pool is embedded in TX/RX-related functions.

\bheading{Conceal shared region and fault handling.} \sysname addresses the loss of security protections from the kernel in the DPDK configuration by concealing the shared memory region. In the case of SEV with VirtIO,
the guest VM's kernel is responsible for formatting, parsing, or \textit{memcopy} raw network packets from/to the shared region. Therefore, the memory content that network applications can touch is guaranteed to always reside in private memory. DPDK, on the one hand, bypasses the traditional kernel networking stack, which consequently disables the security protection provided by the guest VM's kernel layer. Therefore, it also introduces a risk where DPDK network applications may mistakenly access the shared region.
To address this concern, all virtual addresses pointing to shared regions are concealed within the EAL of the DPDK library, hidden from the network application. This ensures that the network application can never directly access the shared region, either intentionally or inadvertently.
In the event of a DPDK crash or faults, both the guest VM OS and the shared region management kernel module ensure that the pages in the shared region are not reused by other programs until the management module zeros out and recycles the region.




\subsection{Efficient I/O Event Handling}

\sysname naturally inherits the efficient optimizations of I/O event handling provided by DPDK, including the use of polling mode to avoid interrupt overhead and the utilization of huge pages to reduce memory lookup costs. Additionally, during DPDK's execution, there is no active and frequent retrieval of timestamps or special instructions that will trigger the VC handler, leading to the avoidance of expensive \texttt{VMEXIT}s. The overhead of the bounce buffer has been significantly reduced by the pre-allocated shadow pool design, ensuring that only one memory copy is performed each time. 

\begin{figure}[t]
\centering
\hspace*{2pt}
\begin{subfigure}[b]{0.32\columnwidth}
\includegraphics[width=\textwidth]{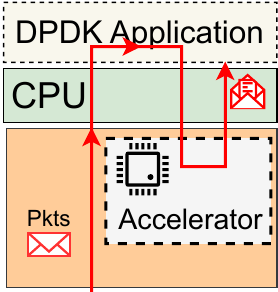}
\caption{\footnotesize Look-aside Mode.}
\label{fig:ipsec_modes:nic_lookaside}
\end{subfigure}
\begin{subfigure}[b]{0.32\columnwidth}
\includegraphics[width=\textwidth]{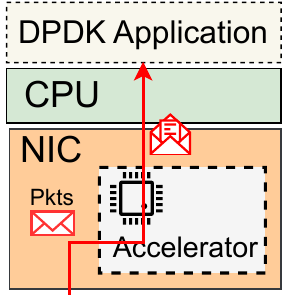}
\caption{\footnotesize Inline Mode.}
\label{fig:ipsec_modes:nic_inline}
\end{subfigure}
\begin{subfigure}[b]{0.32\columnwidth}
\includegraphics[width=\textwidth]{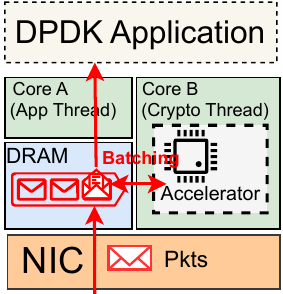}
\caption{\footnotesize Emulated Mode.}
\label{fig:ipsec_modes:cpu_inline}
\end{subfigure}
\vspace{-15pt}
\caption{Three crypto offload methods. }
\vspace{-15pt}
\label{fig:ipsec_modes}
\end{figure}

\subsection{Hardware-accelerated Encryption}
\label{sec:implementation:ipsec}
Although \sysname cannot offload cryptographic operations to the untrusted NIC, it can still leverage CPU crypto instructions to accelerate packet encryption. These crypto instructions remain non-interceptable and secure in the SNP setup, enabling \sysname to securely and efficiently boost packet encryption and decryption.
\sysname supports two crypto offload modes: the look-aside mode and the inline mode. The look-aside mode is an offload mode in which the DPDK application must actively trigger hardware accelerators for decryption (either in NIC or CPU), as shown in \figref{fig:ipsec_modes:nic_lookaside}. \sysname can naturally support such mode by using DPDK's implementation of IPsec by specifying CPU crypto instructions as the accelerator.
Inline mode (\figref{fig:ipsec_modes:nic_inline}) is a mode in which the NIC actively decrypts the packet upon receiving it. To support this mode without trusting the NIC, \sysname introduces a CPU-based emulated inline mode (\figref{fig:ipsec_modes:cpu_inline}).
In the emulated inline mode, the application thread collaborates with a reserved crypto thread to share the TX/RX queues. This crypto thread performs cryptographic tasks for data packets in real-time through polling and batching, efficiently offloading crypto tasks from the application thread. Both modes supported by \sysname keep encryption keys within the vCPU side, thereby eliminating the risk of key leakage. Please note that there may be additional embedded hardware accelerators supported within the SoC that CVMs can leverage to further enhance performance. For instance, Intel QuickAssist Technology~\cite{intel:2023:qat} is a hardware accelerator offering superior performance compared to crypto instructions. It is readily available and integrated into the SoC of Intel Xeon Scalable Processors~\cite{intel:2023:qat} (the CPU series set to support Intel TDX).



%% file: 7-evaluation.tex
\vspace{-5pt}
\section{Evaluation}
\label{sec:evaluation}
\vspace{-5pt}
The testbed is configured the same as listed in \secref{sec:motivation:setup}. Our goal is to demonstrate that \sysname can achieve comparable network performance to the projected optimal solution with TIO. Since TIO devices are not yet available, we specifically focus on comparing the performance of \sysname in SNP VMs with that of a non-TEE VM using DPDK and SR-IOV. 
We believe that even with TIO devices, the performance of the non-TEE VM will still surpass that of a combination of SNP and TIO devices. This superiority can be attributed to the absence of overhead sources due to SNP protection (\eg, bounce buffer), as detailed in Table \ref{tab:factors}. Furthermore, non-TEE VMs are not affected by the PCIe encryption introduced by TIO. Therefore, comparing the performance of \sysname with that of a non-TEE VM, we can provide an estimate of the potential performance gap before the availability of TIO devices. For ease of description, the term "\textbf{performance of non-TEE VM}" mentioned in this section refers to the performance of a non-TEE VM with a SR-IOV VF and DPDK, and the term "\textbf{performance of SNP VM}" refers to the performance an SNP VM with the support of \sysname and the same VF.

\subsection{Simple UDP Echo Server}
\begin{figure}[t]
\centering
\begin{subfigure}[b]{0.9\columnwidth}
\includegraphics[width=\textwidth]{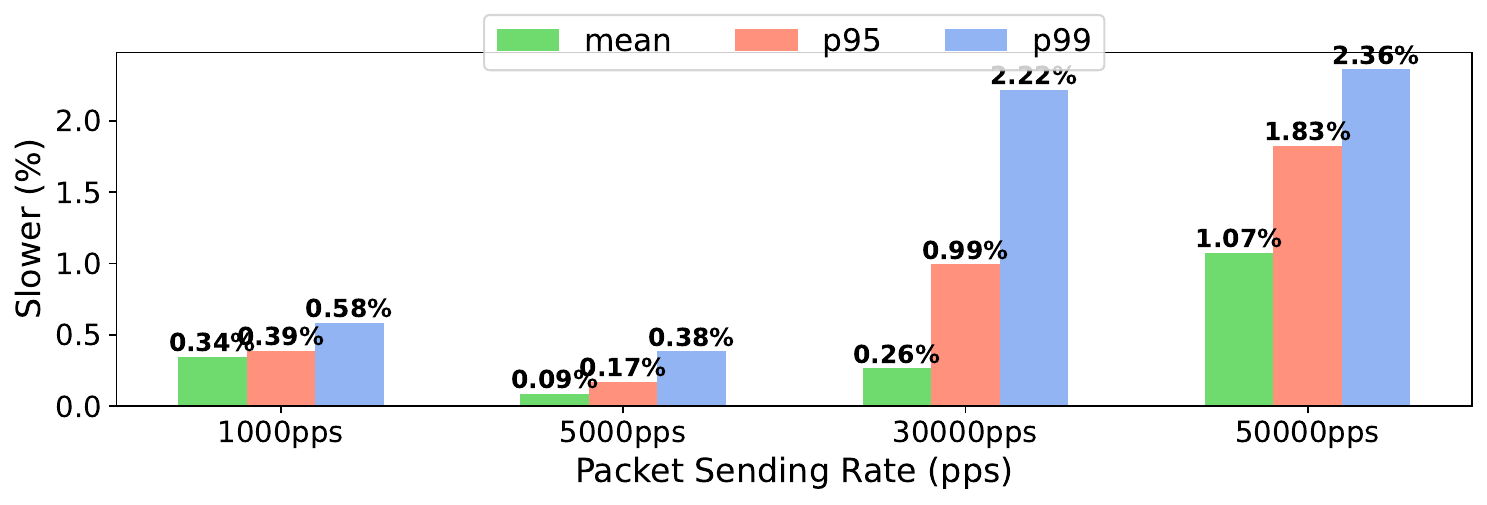}
\vspace{-15pt}
\caption{SNP's percentage of slowdown compared to non-TEE.}
\label{fig:evaluation:echo_snp}
\end{subfigure}
\begin{subfigure}[b]{0.9\columnwidth}
\includegraphics[width=\textwidth]{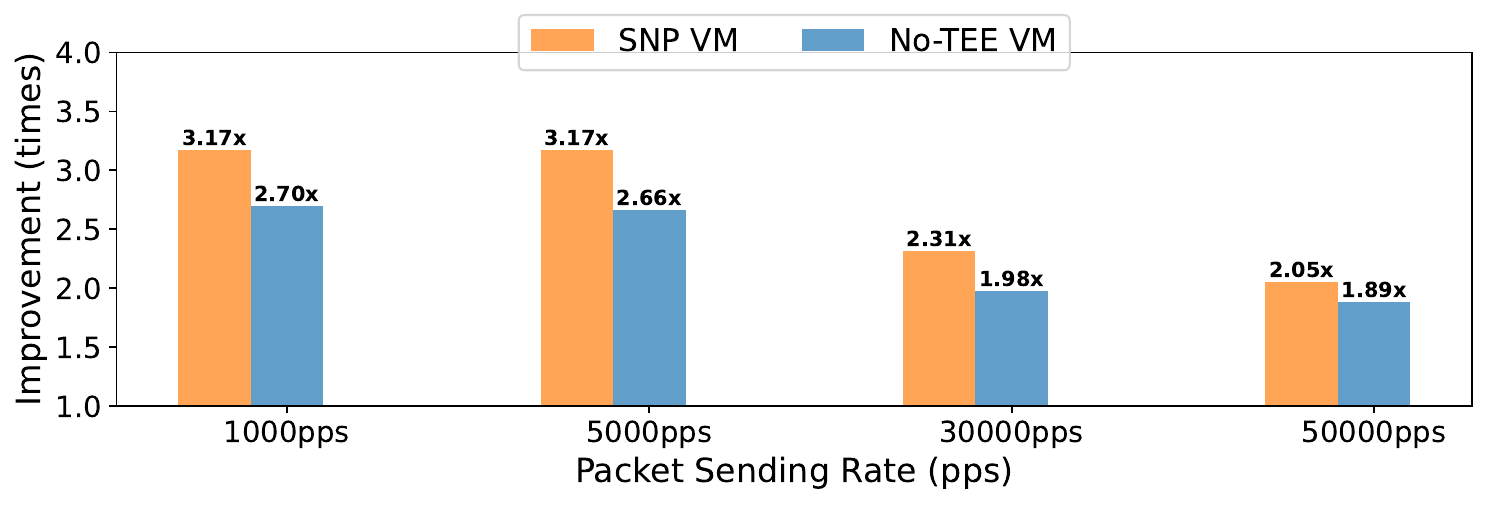}
\vspace{-15pt}
\caption{Improvement compared to SR-IOV only.}
\label{fig:evaluation:echo_improve}
\end{subfigure}
\caption{Tail latency between SNP and non-TEE VMs. }
\label{fig:evaluation:improve_over_sriov}
\vspace{-10pt}
\end{figure}

We first replicated the experiments in \secref{sec:motivation} to provide a tangible understanding and demonstrate the performance differences of \sysname compared to other existing configurations.
\figref{fig:evaluation:echo_snp} presents a comparison of tail latency between \sysname and a non-TEE VM under various packet sending rates. The results indicate that \sysname exhibits an average latency overhead of less than 1\% when compared to the non-TEE VM. Moreover, the p99 tail latency remains within a 2.5\% overhead range. 
In \figref{fig:evaluation:improve_over_sriov}, we show the latency optimization achieved by \sysname in contrast to an SNP VM equipped with SR-IOV alone. 
Across various packet sending rates, \sysname shows an impressive average latency improvement of 2x-3x. This enhancement surpasses the performance optimizations achieved by non-TEE VM configurations, whether utilizing DPDK over SR-IOV or not. These results align with our expectations because non-TEE VMs do not introduce any bounce buffer overhead for both configurations. In contrast, for SNP VMs, compared to the SR-IOV-alone solution, which incurs both bounce buffer allocation and copy overhead, \sysname goes a step further in mitigating the allocation overhead.

\subsection{Generalized Network Testing Tool}
\bheading{Specification.} To comprehensively assess the performance of \sysname under different network traffic, we utilized a well-established DPDK-based benchmark tool called dperf~\cite{baidu:2023:dperf}.
dperf is specifically designed to evaluate network performance across a range of network loads and has garnered significant recognition on GitHub with more than 3.7k stars. We believe that dperf is a comprehensive tool to evaluate \sysname's performance, attributed to its three key advantages: (1) dperf provides a flexible configuration to simulate various network traffic, including packet size, rate, type, number of threads, number of concurrent connections, and connection live time, providing a versatile testing environment.
(2) dperf can simulate heavy network loads. By leveraging the DPDK foundation on both the client and server sides, dperf enables Gigabits per second (Gbps)-level traffic and supports tens of thousands of concurrent connections. (3) dperf provides detailed statistics from different aspects, including packet counts, throughput, and round-trip time (RTT), offering valuable statistics for a thorough analysis.
Benefit from the code compatibility of \sysname, the source code of dperf can be directly compiled without additional source-code modifications.

\bheading{Throughput.}
We first compare the throughput of \sysname and the non-TEE VM. Other than the system's performance, the throughput is inherently constrained by the NIC's performance and the overall network environment. Thus, to present the theoretical maximum bandwidth, we also conducted tests in a non-virtualized environment using Physical Function (PF) directly, marked as ``server (PF)".

\begin{figure}[t]
\centering
\begin{subfigure}[b]{0.45\columnwidth}
\includegraphics[width=\textwidth]{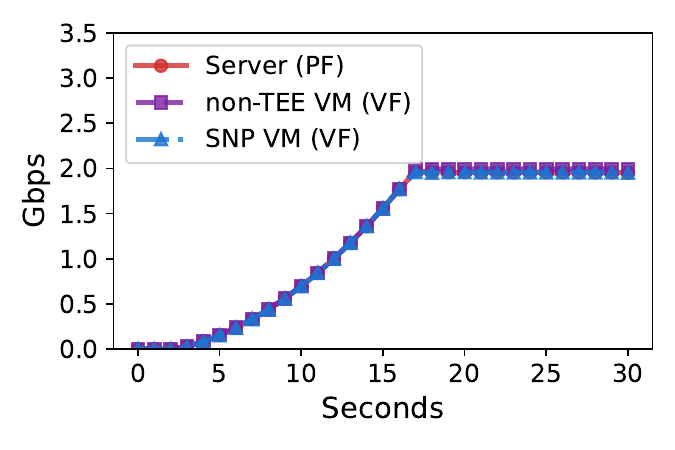}
\vspace{-15pt}
\caption{100 Byte/ packet.}
\label{fig:evaluation:throughput:100}
\end{subfigure}
\begin{subfigure}[b]{0.45\columnwidth}
\includegraphics[width=\textwidth]{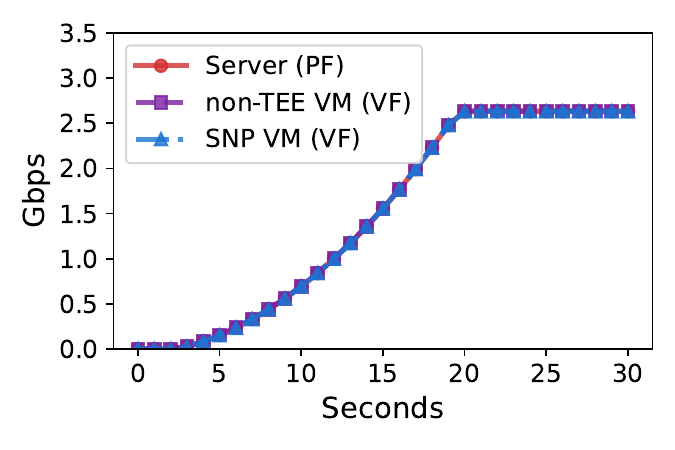}
\vspace{-15pt}
\caption{200 Byte/ packet.}
\label{fig:evaluation:throughput:200}
\end{subfigure}
\begin{subfigure}[b]{0.45\columnwidth}
\includegraphics[width=\textwidth]{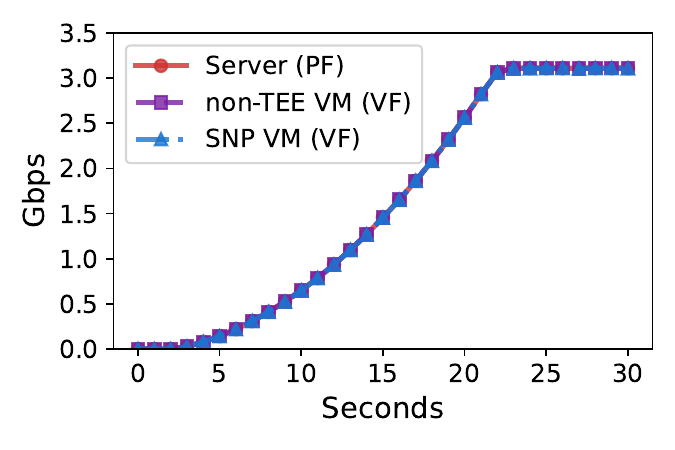}
\vspace{-15pt}
\caption{500 Byte/ packet.}
\label{fig:evaluation:throughput:500}
\end{subfigure}
\begin{subfigure}[b]{0.45\columnwidth}
\includegraphics[width=\textwidth]{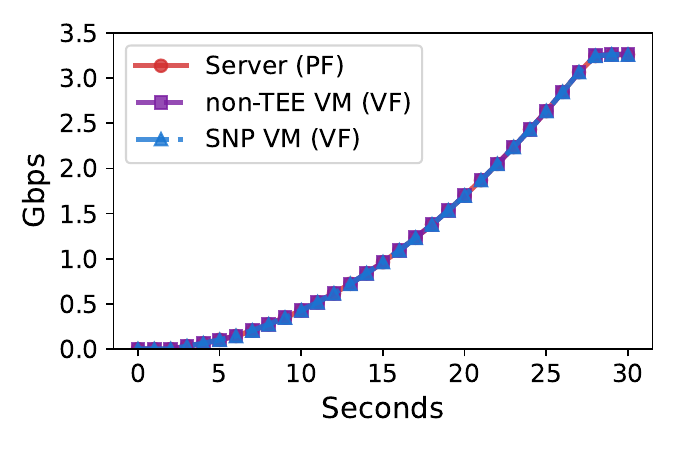}
\vspace{-15pt}
\caption{1000 Byte/ packet.}
\label{fig:evaluation:throughput:1000}
\end{subfigure}
\caption{Throughput under different packet sizes. }
\label{fig:evaluation:throughput}
\vspace{-10pt}
\end{figure}

For evaluating the throughput, we employed UDP as the packet type sent by the client due to its higher speed. The client sent packets at a rate of 1000 packets per second for each connection, utilizing various packet sizes (100 bytes, 200 bytes, 500 bytes, 1000 bytes). To explore the limits, the client incrementally increased the number of concurrent connections per second until packet loss occurred. For every data packet size N bytes, we defined the maximum number of concurrent connections as {\small $\frac{{8 \text{ Gbps}}}{{N \text{byte/ packet} \times 1000 \text{packets/ second}}}$} according to the NIC's capability. The client increased the number of concurrent connections at a rate of 5\% of the maximum number per second. The throughput is calculated by multiplying the number of data packets successfully received by the server per second and the packet size.
As shown in \figref{fig:evaluation:throughput}, the results indicate that the difference in throughput between \sysname and the non-TEE VM was minimal. For various packet sizes (100, 200, 500, 1000 bytes), \sysname's throughput compared to the non-TEE throughput was 98.1\%, 99.9\%, 99.9\%, and 99.8\%, respectively.
Moreover, it is worth noting that our considerations did not account for scheduling. Thus, despite the virtualized environment, the throughput of \sysname closely resembled that of direct execution on the server (``PF" results).

\begin{figure*}[t]
\centering
\begin{subfigure}[b]{0.5\columnwidth}
\includegraphics[width=\textwidth]{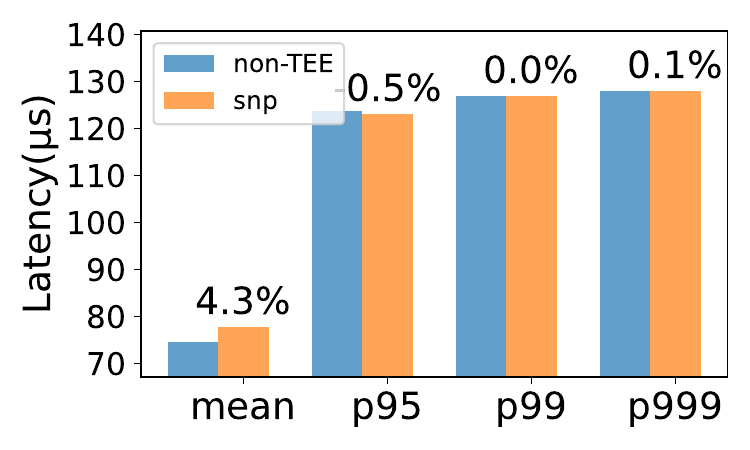}
\vspace{-15pt}
\caption{UDP/1 Byte payload.}
\label{fig:evaluation:tail_1payload}
\end{subfigure}
\begin{subfigure}[b]{0.5\columnwidth}
\includegraphics[width=\textwidth]{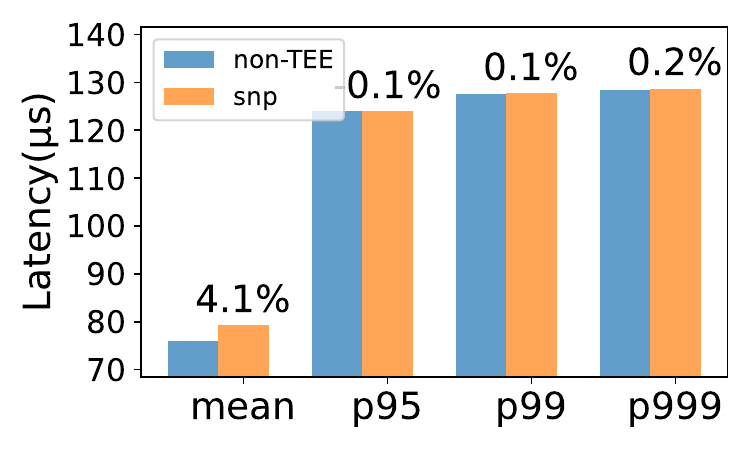}
\vspace{-15pt}
\caption{UDP/100 Byte payload.}
\label{fig:evaluation:tail_100payload}
\end{subfigure}
\begin{subfigure}[b]{0.5\columnwidth}
\includegraphics[width=\textwidth]{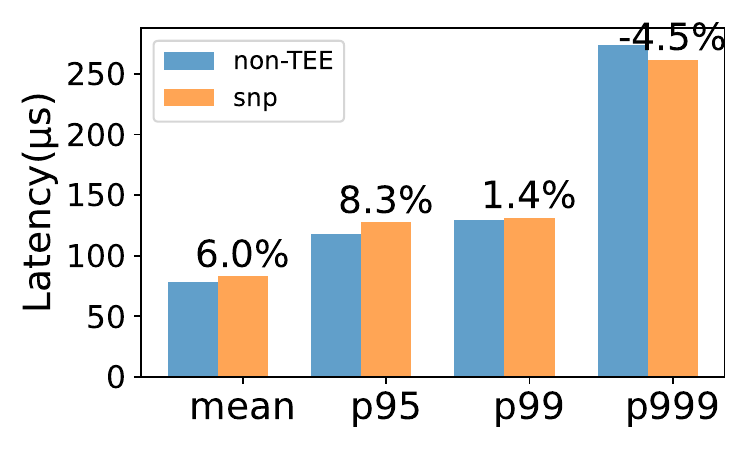}
\vspace{-15pt}
\caption{UDP/500 Byte payload.}
\label{fig:evaluation:tail_500payload}
\end{subfigure}
\begin{subfigure}[b]{0.5\columnwidth}
\includegraphics[width=\textwidth]{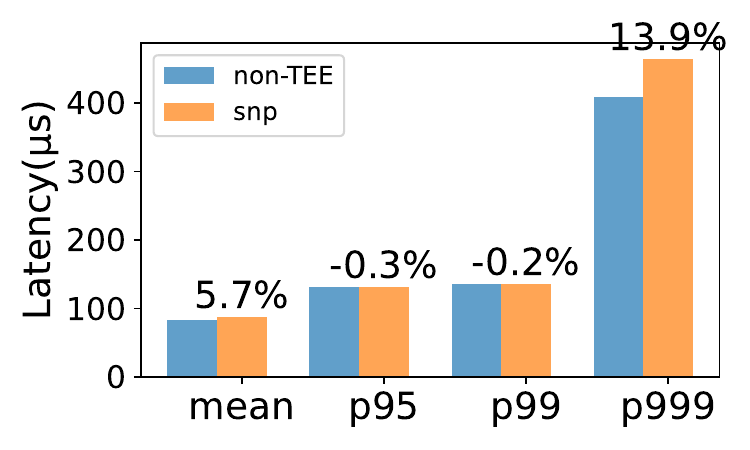}
\vspace{-15pt}
\caption{UDP/1000 Byte payload.}
\label{fig:evaluation:tail_1000payload}
\end{subfigure}
\begin{subfigure}[b]{0.5\columnwidth}
\includegraphics[width=\textwidth]{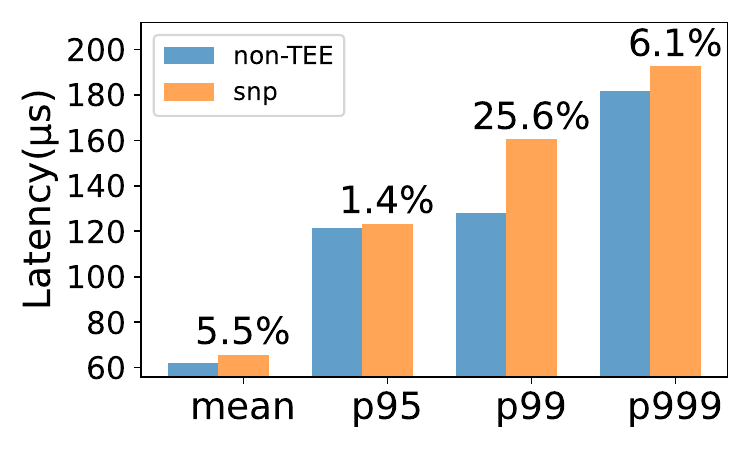}
\vspace{-15pt}
\caption{TCP/1 Byte payload.}
\label{fig:evaluation:tcp:tail_1payload}
\end{subfigure}
\begin{subfigure}[b]{0.5\columnwidth}
\includegraphics[width=\textwidth]{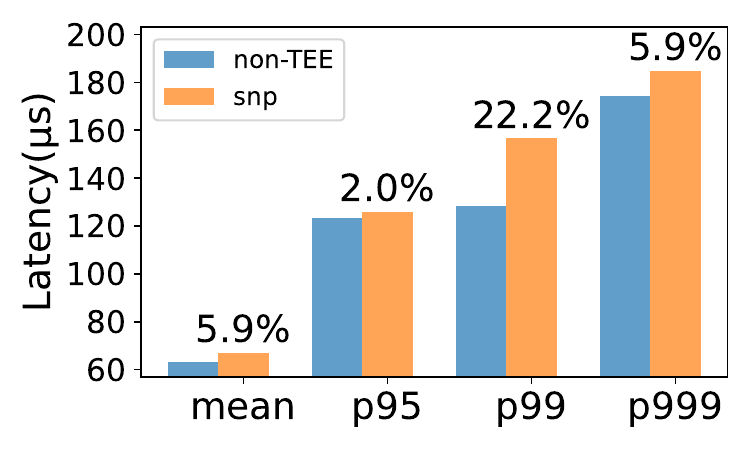}
\vspace{-15pt}
\caption{TCP/100 Byte payload.}
\label{fig:evaluation:tcp:tail_100payload}
\end{subfigure}
\begin{subfigure}[b]{0.5\columnwidth}
\includegraphics[width=\textwidth]{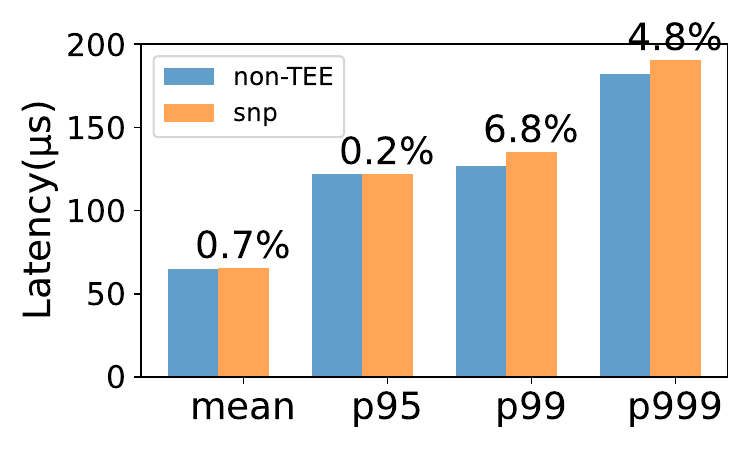}
\vspace{-15pt}
\caption{TCP/500 Byte payload.}
\label{fig:evaluation:tcp:tail_500payload}
\end{subfigure}
\begin{subfigure}[b]{0.5\columnwidth}
\includegraphics[width=\textwidth]{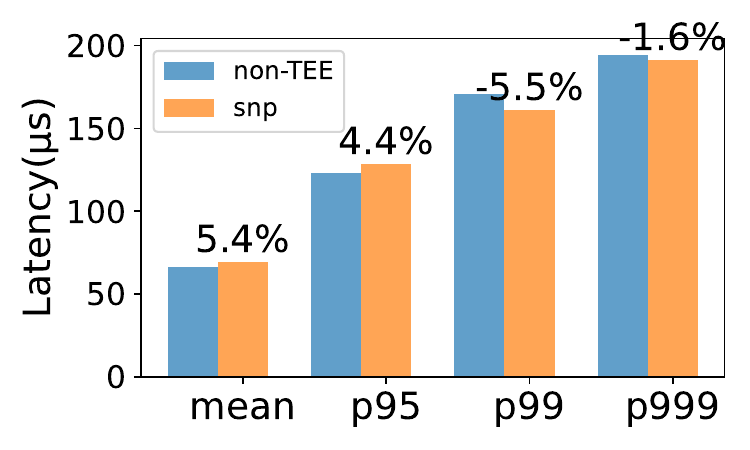}
\vspace{-15pt}
\caption{TCP/1000 Byte payload.}
\label{fig:evaluation:tcp:tail_1000payload}
\end{subfigure}
\caption{Mean and tail latency for UDP and TCP workload. }
\vspace{-10pt}
\label{fig:latency_tcp_udp}
\end{figure*}


\bheading{Latency.}
Latency represents the actual time taken for data processing on the server side. To evaluate latency, we conducted tests under various stress levels, measuring the round-trip time on the client side, which is the time between the client initiating a new connection and receiving the first server-side response.
For UDP configuration, the client gradually increases the number of new connections per second with a payload size of N until it reaches 20,000 (approximately 1/3 of the total port number 65535), and then maintains this level of connections for a stable period of 30 seconds. Under the TCP configuration, the client follows the three-way handshake principle to establish connections, and the server carries data of size N during the third handshake. After one round of communication, each connection is closed.
\figref{fig:latency_tcp_udp} shows that \sysname's average latency is consistently less than 6\% slower than the non-TEE latency across different packet sizes. Additionally, a comparison between UDP and TCP packets reveals that TCP exhibits relatively noticeable tail latency. In cases with the 1-byte payload and 100-byte payload, the tail latency difference at 5\% level is approximately 25\%.

\begin{figure}[t]
\centering
\begin{subfigure}[b]{0.43\columnwidth}
\includegraphics[width=\textwidth]{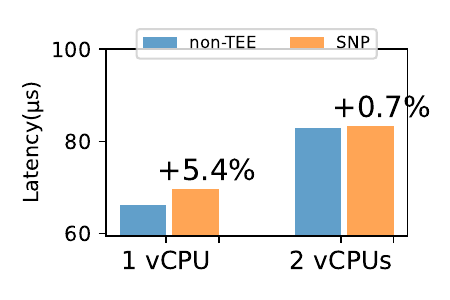}
\vspace{-18pt}
\caption{Mean.}
\label{fig:evaluation:2cpu_mean}
\end{subfigure}
\begin{subfigure}[b]{0.43\columnwidth}
\includegraphics[width=\textwidth]{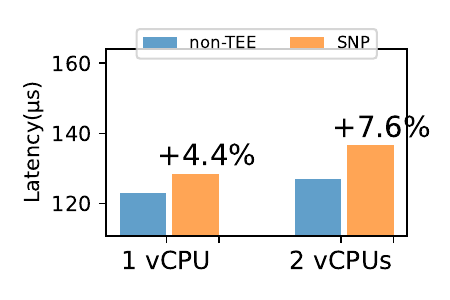}
\vspace{-18pt}
\caption{Tail latency/p95.}
\label{fig:evaluation:2cpu_p95}
\end{subfigure}
\begin{subfigure}[b]{0.43\columnwidth}
\includegraphics[width=\textwidth]{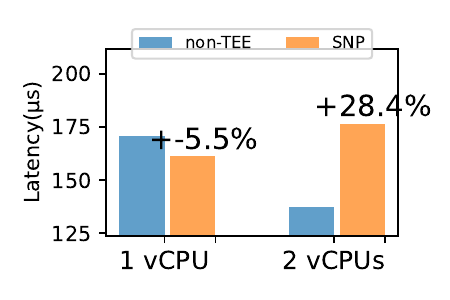}
\vspace{-18pt}
\caption{Tail latency/p99.}
\label{fig:evaluation:2cpu_p99}
\end{subfigure}
\begin{subfigure}[b]{0.43\columnwidth}
\includegraphics[width=\textwidth]{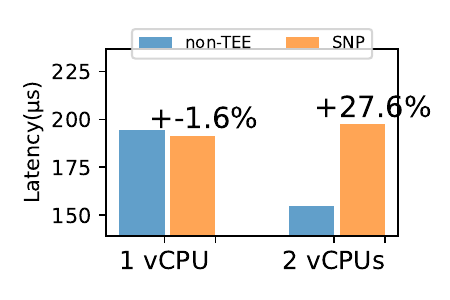}
\vspace{-18pt}
\caption{Tail latency/p999.}
\label{fig:evaluation:2cpu_p999}
\end{subfigure}
\vspace{-5pt}
\caption{Comparison between 1 vCPU and 2 vCPUs. }
\label{fig:2cpus}
\end{figure}

\bheading{Multi-thread.} We tested the performance of \sysname on multiple threads using the RSS (Receive Side Scaling) feature. By configuring RSS, all receiving network packets are shared across multiple threads and processor cores. In the same latency testing environment, we measured the latency of UDP connections with a payload size of 1000 bytes under scenarios with one or two threads. \figref{fig:2cpus} illustrates that using two threads resulted in a slight increase in client-side latency, potentially attributable to the additional routing time incurred by RSS. However, regardless of using one or two threads, \sysname's average latency consistently stays within 6\% of that of non-TEE setup.


\subsection{IPsec Performance}
To evaluate \sysname's performance with IPsec, we enhanced the dperf to support IPsec data streams, which serves as a useful tool to measure encrypted traffic in CVM environment. For simplicity, we hardcoded the IPsec keys on both the server and client sides, focusing on analyzing the performance impact of the encrypted data stream from end to end.
The IPsec connection is configured to use a transport mode, where source and destination addresses are not encrypted for routing purposes. The IPsec decryption method was configured to use the 128-bit AES algorithm in GCM mode, with each packet carrying 1000-byte data. It is essential to note that due to hardware limitations with the network card, the VF driver used in our experiments did not support IPsec offload inline mode, even in a non-virtualized environment. As a result, the results for IPsec offload inline mode are missing. Moreover, the performance of hardware-accelerated encryption could be highly dependent on the CPU model (\eg, CPU frequency, and CPU architecture), or the accelerators used by the NIC. Thus, we used a CPU look-aside mode running in a non-virtualized environment with PF directly as the reference for upper bound (marked as ``Server (PF)"), to evaluate the performance of \sysname with encrypted traffic. 


\begin{figure}[t]
\hspace*{17pt}
\includegraphics[width=0.82\columnwidth]{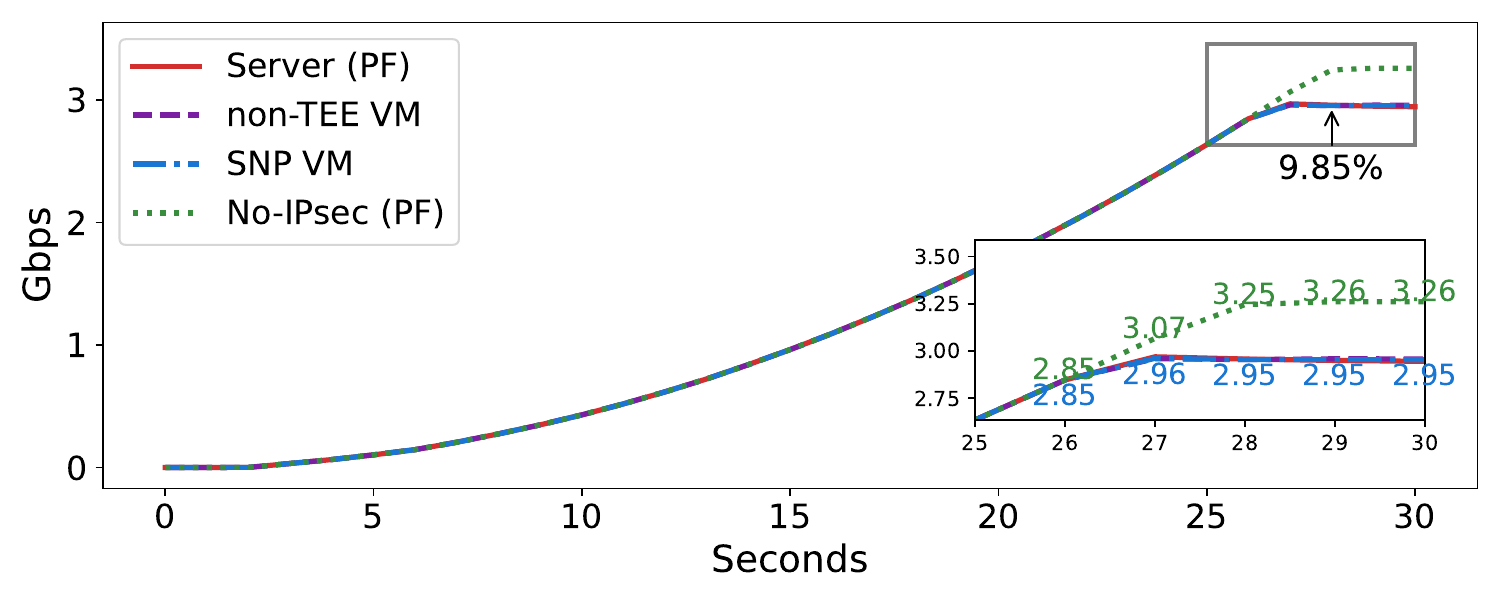}
\vspace{-5pt}
\caption{Throughput when enabling IPsec. }
\label{fig:ipsec_throughput}
\vspace{-10pt}
\end{figure}

\begin{figure}[t]
\hspace*{13pt}
\includegraphics[width=0.85\columnwidth]{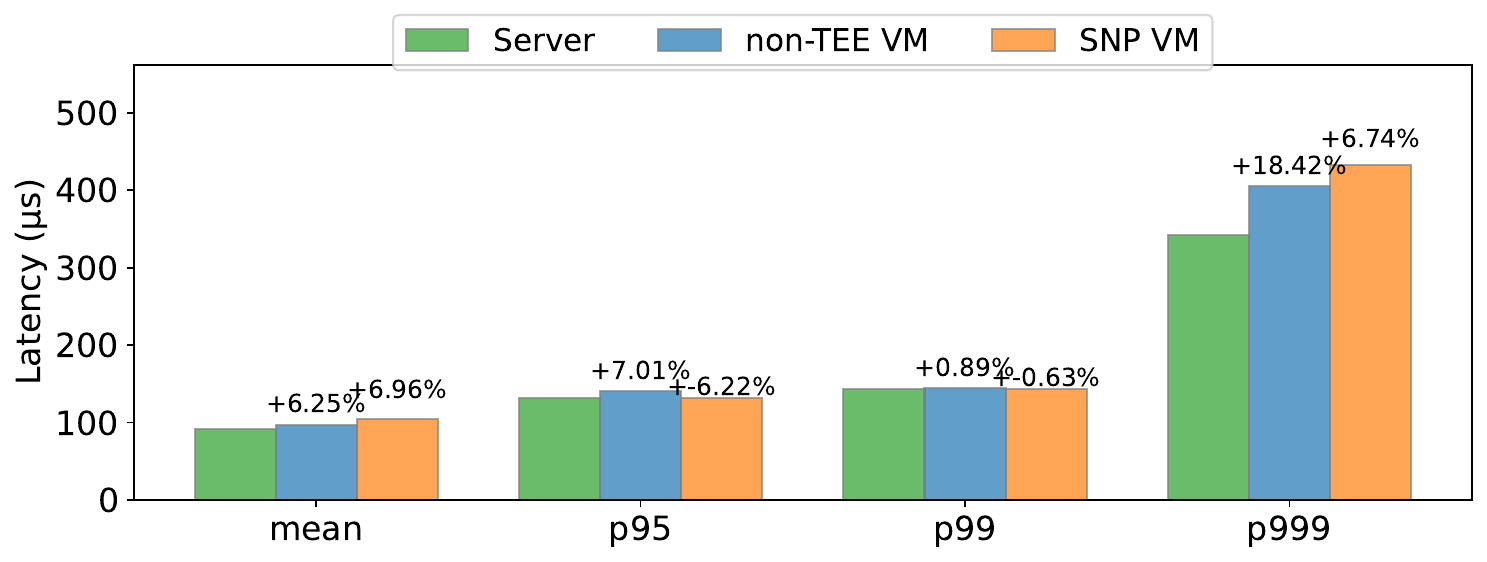}
\vspace{-5pt}
\caption{IPsec latency. }
\label{fig:ipsec_latency}
\vspace{-5pt}
\end{figure}

\bheading{CPU look-aside mode.}
The throughput and latency tests were set up similarly to the tests without IPsec. Regarding throughput, as shown in \figref{fig:ipsec_throughput}, \sysname's throughput with IPsec enabled was nearly equivalent to that of the non-TEE VM or the native non-virtualized environment. Compared to the non-TEE VM, \sysname achieved 99.8\% of its throughput. However,  IPsec does have an impact on throughput, resulting in a decrease of 9.85\% when compared to non-IPsec traffic. Regarding latency, \figref{fig:ipsec_latency} shows the difference in percentage between non-TEE VM and the server, and between \sysname and non-TEE VM. On average, \sysname exhibited a latency 6.96\% higher than the non-TEE VM. 

\begin{figure}[t]
\hspace*{10pt}
\includegraphics[width=0.87\columnwidth]{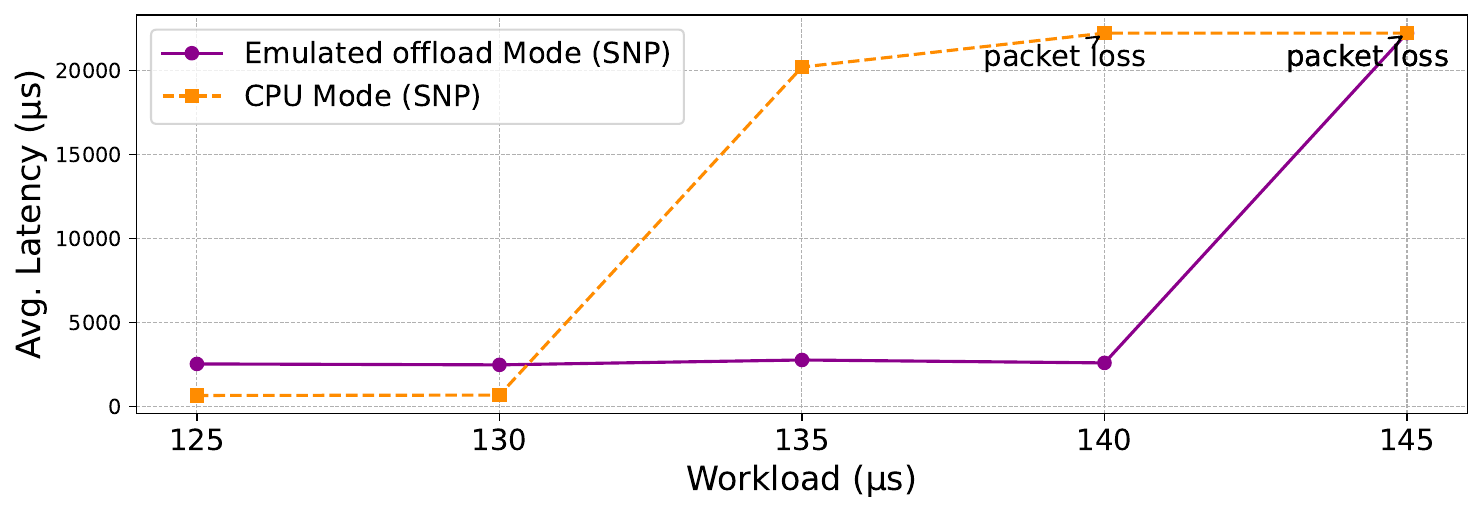}
\vspace{-5pt}
\caption{Performance of emulated-inline mode. }
\label{fig:emulation_mode}
\vspace{-10pt}
\end{figure}

\begin{figure}[t]
\centering
\begin{subfigure}[b]{0.49\columnwidth}
\includegraphics[width=\textwidth]{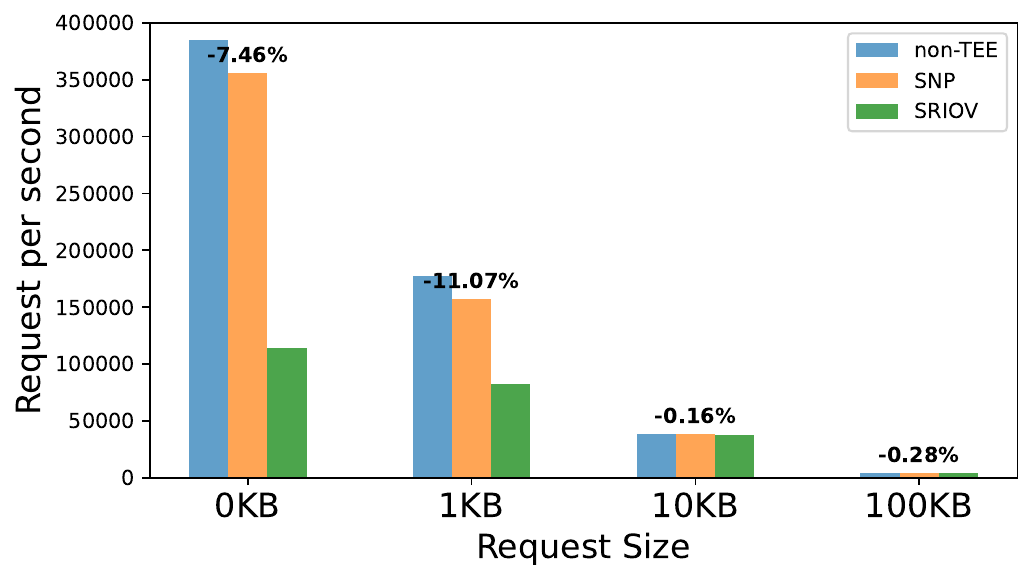}
\vspace{-15pt}
\caption{RPS Comparison.}
\label{fig:nginx:rps}
\end{subfigure}
\begin{subfigure}[b]{0.49\columnwidth}
\includegraphics[width=\textwidth]{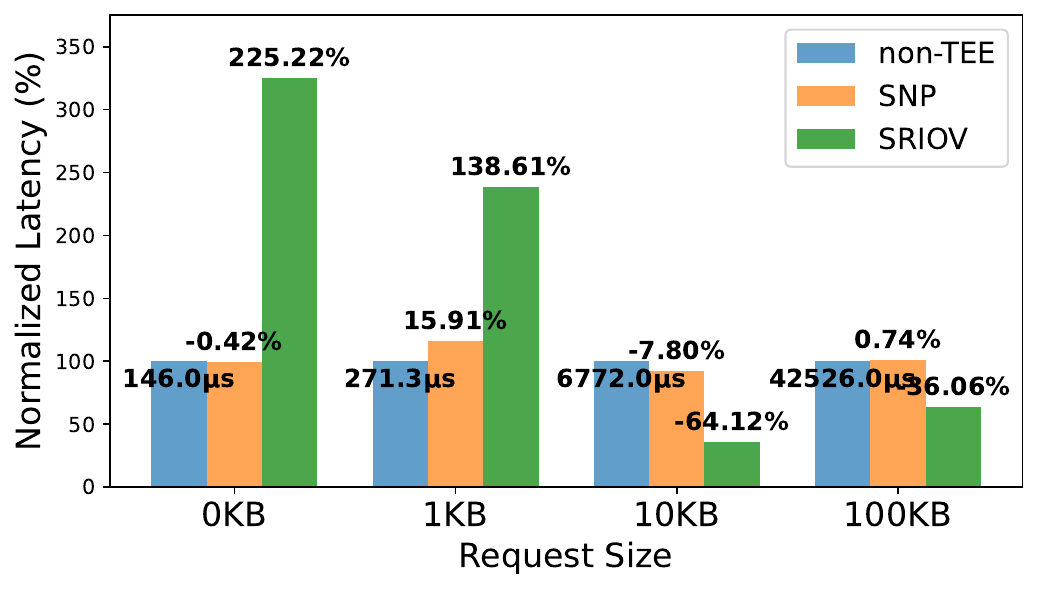}
\vspace{-15pt}
\caption{Latency Comparison.}
\label{fig:nginx:latency}
\end{subfigure}
\vspace{-5pt}
\caption{Nginx Performance. }
\label{fig:nginx}
\vspace{-10pt}
\end{figure}

\bheading{CPU-enabled emulated-inline Mode.}
To evaluate the emulated inline mode, we introduced additional simulated per-packet processing time (workload after decryption) on the server side and measured latency at different workloads. We opted for this test because emulated inline mode requires reserving an extra vCPU for polling and batching. However, in real network applications, expanding one more application thread on this extra vCPU should be a wiser choice, as it eliminates the need for handling shared TX/RX queues or dispatching issues by introducing the crypto thread. Thus, an emulated inline mode may only be meaningful when the network application supports only one thread and suffers from CPU pressure. As expected, the experimental results with 5000pps show that the emulated mode results in higher latency due to the overhead of TX/RX queue synchronization. However, the emulated mode can alleviate pressure on the application thread. As shown in \figref{fig:emulation_mode}, when the application thread takes more than 135 milliseconds to process per IPsec packet, the thread experiences a 10x increase in latency due to higher pressure. In contrast, under emulated mode, significant latency and packet loss occur only at a workload of 145 milliseconds on the server side.

\subsection{Real World DPDK Applications}
\label{sec:evaluation:apps}
To demonstrate the code compatibility of \sysname and showcase its performance in real-world applications, we evaluated the performance of nginx~\cite{nginx:2023:nginx} server based on F-Stack~\cite{fstack:2023:fstack} (\texttt{Commit:dd27d06}). 
F-stack is an open-sourced network framework based on DPDK, providing POSIX API (Socket, Epoll, Kqueue), user-space TCP/IP stack (port FreeBSD), programming SDK (Coroutine), and application interfaces (nginx), to assist in helping network applications benefit from DPDK.
We focus on comparing the performance of F-stack-provided nginx on \sysname with non-TEE VM. As a control, we also show the performance of an ordinary version of nginx obtained from \texttt{apt-get} running in a non-TEE VM with SR-IOV.
We measured the requests-per-second (RPS) and latency following an official guidance~\cite{nginx:2017:nginx-perf}. 
The client side ran a HTTP benchmarking tool, called wrk~\cite{wrk:2023:github}. We collected each metric against various requested file sizes. For the RPS test, we ran $12$ independent wrk to gather the maximum RPS. Each instance opened $50$ HTTP connections per minute. For the latency test, we ran $1$ wrk for 5 minutes to get stable results.

\figref{fig:nginx:rps} shows the total RPS. While there is a performance gap of $11.07\%$ for the $1$KB test, this gap shrinks to negligible as the request size increases.
\figref{fig:nginx:latency} shows the average latency. When the request size exceeds 10KB, the original nginx shows better performance, which could be attributed to some specific optimizations in the original nginx. Conversely, the nginx provided by F-stack might not consider such optimizations for large request sizes. As a result, the primary focus for larger request dimensions is the performance variance between the identical F-stack nginx in SNP VM (\sysname) and non-TEE VM.
Note that the higher performance difference shown in these results is a combination of the performance difference for both program execution and network operations, which highly depends on the implementation of the program. Even TIO-based solutions cannot mitigate the overhead due to program execution in a SNP VM. These F-stack experiments also indicate that \sysname can be applied to complex existing network frameworks.

%% file: 8-discussion.tex
\section{{\fontsize{12pt}{1}\selectfont Comparison of \sysname and TIO Solutions}}
\label{sec:discussion}
\subsection{Security Comparison}
\label{sec:discussion:security}
Due to distinct threat models, \sysname and TIO solutions attain varying security levels and TCB sizes.

\bheading{\sysname: end-to-end security.}
\sysname shares the same threat model and accomplishes the same end-to-end security level as the default VirtIO solution. They both exclusively rely on the CPU side, thereby necessitating the employment of software-based encryption for ensuring comprehensive I/O security. 
Various software-based solutions can be utilized to ensure end-to-end security, such as IPsec at the network layer or SSL/TLS at the application layer. Considering the performance of SSL/TLS may vary based on the application being used, we use IPsec in this paper to evaluate the potential throughput when software-based encryption is enabled. Furthermore, it is important to mention that vulnerable DPDK applications or device drivers are beyond the scope of this paper, as SNP's threat model assumes that all software components within SNP VMs are trustworthy. Some I/O security studies related to identify internal bugs are included in \appref{appendix:io}.
  
\bheading{TIO: VM-to-NIC security.}
TIO solutions achieve a different VM-to-NIC security. The different threat model adopted by TIO, where the device and VM both reside within the TCB, allows them to operate within the private memory. This, together with encryption on the PCIe buses, ensures the security of VM-to-NIC communication. 
However, as the paper points out, such link protection over the PCIe bus is redundant, as network packets are supposed to be protocol encrypted in TEE model.
It's worth noting that the TIO reduces certain attack vectors during I/O communication. For instance, the driver interfaces now reside completely in private memory, mitigating attacks that attempt to explore vulnerable driver implementations~\cite{hetzelt:2021:via}. A detailed discussion of existing SEV attacks' impact on these two I/O models is provided in \appref{appendix:attacks}.

\bheading{TCB size.} Compared to TIO solutions which should significantly increase TCB size, \sysname slightly increases the TCB size compared to the original software stack. \sysname introduced approximately 2K lines of code modifications to the DPDK library, and the shared region management kernel module added 0.5K lines of code. Additionally, we added 1.2K lines of code to dperf to support IPsec benchmarking. Apart from these changes, we made less than 5 lines of code changes to the F-Stack source code, primarily focusing on fixing the inconsistent kernel headers, to ensure successful compilation with the kernel version running in our VMs. 





\subsection{Performance Comparison}
\sysname can only benefit DPDK-compatible applications and has extra memory overhead compared to TIO solutions.

\bheading{Network application without DPDK.} Although \sysname can achieve the same-level performance compared to the optimal solution, where both DPDK and TIO are utilized, 
network applications without DPDK support cannot directly benefit from \sysname. On the contrary, with TIO support, network applications using POSIX APIs can also potentially benefit from the overhead mitigated by the TIO device. 
Luckily, \sysname can work together with some software-based network frameworks, like F-stack~\cite{fstack:2023:fstack}, to alleviate SNP-specific overhead for applications using POSIX APIs. These frameworks embed DPDK to provide POSIX APIs or other network socket interfaces, such as user-space TCP/IP stacks, to help network applications benefit from DPDK.

\bheading{Additional Memory Overhead.} One limitation of \sysname is the additional memory overhead. To avoid the extra allocation delay during runtime, \sysname allocates all data structures required by the shadow memory pool design during DPDK initialization. 
By default, this setting approximately doubles the memory pool size used, as we configure the shared memory pool to be the same size as the shadow memory pool.
Compared to VMs with memory in the GB range, this overhead typically falls within hundreds of MBs. For instance, in the evaluation's dperf setup, when using the default configuration with one NIC port, an additional 65,456 mbufs are allocated, each having a size of 2,176 bytes, resulting in an approximate additional memory usage of 136 MBs.
The overhead in the F-stack example varies based on configuration, as users have the flexibility to set up F-stack with different numbers of queues. By default, 8,192 mbufs are allocated for each queue, each having a size of 2,176 bytes, which leads to an additional memory overhead of approximately 17 MBs.







%% file: 9-related.tex
\section{Related Work}
\label{sec:related}

\bheading{TEE performance optimizations.}
During the development of our paper, we became aware of another project~\cite{li:2023:analysis} that is also investigating network performance in CVMs. Despite the similarities in the research area, there are three key differences between our work and theirs:
(1) Focus on overhead identification: Our work focuses primarily on conducting a comprehensive analysis of the impact of security measures implemented in the design of CVMs. In contrast, their approach centers around breaking down the time spent on different components (host side and VM side) to identify specific overheads, such as the time spent during \texttt{VMEXIT}. 
(2) Research vision: Our project's main objective revolves around enhancing future high-performance I/O capabilities with support for SR-IOV and DPDK on the latest SNP machines. Conversely, their emphasis is placed on improving the performance of QEMU-based emulated I/O devices on SEV-ES machines.
(3) Optimization methodology: They employ strategies to mitigate the overhead caused by the bounce buffer itself. Their methods enable zero-copy properties, such as directly placing encrypted packages in shared memory, primarily benefiting from in-kernel TLS layer modifications and packet pre-processing. These modifications result in slightly altered software behavior within conventional CVMs, where software never writes directly to the shared memory except for memory copying.
In contrast, our approach strictly follows the constrained interfaces as what CVMs now do to ensure security and focuses on eliminating all factors that lead to poor performance due to SNP protection . 

 
Other related TEE network performance optimizations~\cite{arnautov:2016:scone,orenbach:2017:eleos,priebe:2019:sgx-lkl} primarily target enclave-based TEE (Intel SGX~\cite{costan:2016:intel}) and involve using extra I/O threads to achieve exitless I/O communication with SGX. Rkt-I/O~\cite{thalheim:2021:rktio} also utilizes DPDK to enhance network performance by embedding it into LibOS as a hidden network driver. However, a significant distinction between our work and theirs is that our approach fully utilizes DPDK without the LibOS layer, motivating us to provide a security-oriented set of interfaces distinct from theirs, while ensuring end-to-end IPsec and maintaining code compatibility.
Remarkably, our research also reveals the insight that existing SNP VMs, with the support of \sysname, can also achieve the same level of performance as future TIO solutions. 

%% file: 10-conclusion.tex
\vspace{-5pt}
\section{Conclusion}
\label{sec:conclusion}
\vspace{-5pt}
In this paper, we propose a DPDK-based solution called \sysname that achieves network performance close to that of the projected optimal network solution with TIO support. \sysname relies solely on CPU trust and allows existing confidential VMs to benefit from comparable network I/O performance until TIO is market-ready and thoroughly examined.

%% file: 0-appendix.tex
\section{Analysis of Exposed Metadata}
\label{appendix:metadata}

Tables \ref{table:txdesc} and \ref{table:rxdesc} show the complete list of fields in the TX and RX descriptor rings. The descriptors are in the same format as used by the drivers in the operating system. The "Read/ Write" column indicates how the VM side will operate on this metadata. The write-only fields are trivially secure to be exposed in the shared memory, as the VM will only write to this field. Regarding read-only fields, most are read only once and subsequently copied to private memory. The only two exceptions are the ``status" in the TX ring and ``status\_error" in the RX ring, which are read until the descriptor reaches a ``ready'' state (or an error). Once the descriptor is ready, this field does not need to be read again, making it safe to be placed in shared memory.

\begin{table}[h]
\scalebox{0.9}{
\begin{tabular}{@{}lll@{}}
\toprule
\hline
\multicolumn{3}{c}{TX Descriptor Rings}                                                                                                 \\ \hline
\multicolumn{1}{l|}{\textbf{Field}}          & \multicolumn{1}{l|}{\textbf{Read / Write}} & \textbf{Notes}                                                                                                     \\\hline
\multicolumn{1}{l|}{Address}        & \multicolumn{1}{l|}{Write}        &                                                                                                           \\
\hline
\multicolumn{1}{l|}{cmd\_type\_len} & \multicolumn{1}{l|}{Write}        & Length and various flags.                                                                                  \\
\hline
\multicolumn{1}{l|}{olinfo\_status} & \multicolumn{1}{l|}{Write}        & \begin{tabular}[c]{@{}l@{}}Various flags for offloading \\ features.\end{tabular}                          \\
\hline
\multicolumn{1}{l|}{status}         & \multicolumn{1}{l|}{Read}         & \begin{tabular}[c]{@{}l@{}}Only read until status is ``free'',\\ making it safe to read from.\end{tabular}
\end{tabular}}
\caption{Fields in the TX Descriptor Rings}
\label{table:txdesc}
\end{table}

\begin{table}[h]
\scalebox{0.85}{
\begin{tabular}{@{}lll@{}}
\toprule
\hline
\multicolumn{3}{c}{RX Descriptor Rings}                                                                                                                                                     \\ \hline
\multicolumn{1}{l|}{\textbf{Field}}          & \multicolumn{1}{l|}{\textbf{Read / Write}} & \textbf{Notes}                                                                                                             \\ \hline
\multicolumn{1}{l|}{Packet Address} & \multicolumn{1}{l|}{Write}        &                                                                                                                   \\ \hline
\multicolumn{1}{l|}{Header Address} & \multicolumn{1}{l|}{Write}        & Length and various flags.                                                                                          \\ \hline
\multicolumn{1}{l|}{Packet Info}    & \multicolumn{1}{l|}{Read}         & Contains type of packet.                                                                                           \\ \hline
\multicolumn{1}{l|}{RSS}            & \multicolumn{1}{l|}{Read}         & \begin{tabular}[c]{@{}l@{}}Value used for hashing and scaling,\\ only read once and stored into mbuf.\end{tabular} \\ \hline
\multicolumn{1}{l|}{Status\_Error}  & \multicolumn{1}{l|}{Read}         & \begin{tabular}[c]{@{}l@{}}Only read until packet is ready or \\error.\end{tabular}                               \\ \hline
\multicolumn{1}{l|}{VLAN Tag}       & \multicolumn{1}{l|}{Read}         & \begin{tabular}[c]{@{}l@{}}VLAN tag control identifier, only \\read once and stored into mbuf.\end{tabular}        \\ \hline
\multicolumn{1}{l|}{Length}         & \multicolumn{1}{l|}{Read}         & \begin{tabular}[c]{@{}l@{}}Packet length, only read once \\ and stored into mbuf.\end{tabular}                     
\end{tabular}}
\caption{Fields in the RX Descriptor Rings}
\label{table:rxdesc}
\end{table}

\section{Attacks against SEV}
\label{appendix:attacks}
SEV series (SEV/SEV-ES/SEV-SNP) has been subjected to numerous existing attacks. 
Luckily, SEV-SNP addresses most those attacks and here we outline several key attack categories against SEV series.

\bheading{Unencrypted registers [SEV-only].} In SEV, the register values backed up in the VM control block (VMCB) during VMEXIT are not encrypted, giving attackers~\cite{hetzelt:2017:security} ~\cite{werner:2019:severest} a time window to inspect and manipulate the register values. Starting from SEV-ES, register values are encrypted and stored in a VM save area (VMSA), and thus such attacks are mitigated.

\bheading{Weak encryption [SEV/ES].} In the early production lines of SEV (affecting a range of EPYC 7xx1 and 7xx2 processors), the tweak function used in memory encryption was not entirely random. This allowed attackers to perform reverse engineering on the tweak function~\cite{buhren:2017:fault,du:2017:sevUnsecure,wilke:2020:sevurity}, enabling them to eliminate the randomness introduced by the tweak function and copy ciphertext to other memory addresses. AMD introduced SEV-SNP starting from the Zen3 architecture's Milan series (EPYC 7xx3), and the corresponding bugs are fixed and will not impact SEV-SNP.

\bheading{Unauthorized encryption [SEV/ES].} Attackers can manipulate VM;s Address Space ID (ASID) to make the VM use someone other's memory encryption key for decryption. By controlling a malicious VM, the attacker~\cite{li:2020:crossline} can attempt to access the victim's memory pages, creating a brief attack window to decrypt the victim's memory during page table walk. SEV-SNP mitigates such attacks by adding memory integrity protection, where all write operations are checked to verify if the executor is the legitimate owner. 

\bheading{Unprotected memory mapping [SEV/ES].} In SEV and SEV-ES, attackers can tamper with nested page tables to remap guest physical addresses to other memory pages, leading to incorrect data pages or instruction pages being accessed by the VM~\cite{morbitzer:2018:severed,morbitzer:2021:severity_new, morbitzer:2019:extract_new}. The memory integrity protection provided by SNP can also mitigate such attacks since the mapping information between guest physical addresses to system physical addresses are also recorded and checked.

\bheading{Fault injection [SEV/ES/SNP].} The fault injection attack~\cite{buhren:2021:one} is a potential threat that could impact SNP and requires physical access to the CPU chip. The attacker can introduce hardware faults during the CPU boot process to bypassing certain security checks and enabling the AMD secure processor to execute flawed firmware. Such attacks can affect both TIO solutions and the current I/O solution, but should be fixable by firmware and are not relevant to our paper, as we only care about I/O operations when the server machine has fully booted.

\bheading{Side-channel and controlled-channel attacks.} Even though side-channel and controlled-channel attacks are beyond the scope of the threat model of most commercial TEE designs, such as AMD SEV~\cite{kaplan:2016:sevWpaper} and Intel SGX~\cite{costan:2016:intel}, 
existing side-channel attacks have been shown to be capable of leaking information from an SNP-protected VM to varying degrees. Apart from traditional cache side-channels that can work even without the strong threat model of a TEE, controlled-channel attacks are more powerful and fine-grained side-channels specifically designed for TEE setups.
Controlled-channel attacks gain more precise side-channel information by explicitly controlling some of the resources used by the TEE. For instance, the Page-level controlled-channel allows attackers to infer the next branch or intercept the time when a TEE instance tries to access a specific memory page by controlling the "present" bit in the page table~\cite{wilke:2020:sevurity,morbitzer:2021:severity_new,wilke:2021:undeserved,li:2021:tlb}. The interrupt-based controlled-channel can enable attackers to single-step TEE instances to get instruction-level attack window~\cite{li:2021:cipherleaks,van:2017:sgx}.

Moreover, in the context of TEE setups, attackers can access additional side-channel information, such as inferring VM instructions based on CPU power consumption~\cite{wang:2023:pwrleak} or deducing VM behavior using Performance Monitoring Counters (PMC)~\cite{li:2022:systematic}.
One side-channel attack that still remains effective in SNP is the ciphertext side-channel. Because of most commercial TEE encrypt the memory encryption without introducing freshness, enabling attackers to infer plaintext by observing ciphertext. Users can prevent such leakage by modifying their source code running inside SNP VM in a ciphertext-side-channel-resident way~\cite{amd:2022:ciphertext}. All side-channel attacks can potentially be utilized in both TIO and the current I/O solutions, as TIO and SNP do not introduce additional protection against side-channel attacks.

\section{TEE I/O security}
\label{appendix:io}
Hetzelt~\etal~\cite{hetzelt:2021:via} develop a dynamic fuzzing tool for testing device interfaces in confidential VM, identifying 50 bugs in various device drivers. Their work holds significant value in real-world TEE setups, as all code running inside the confidential VM is considered trusted within its thread model. 
Therefore, driver bugs, such as buffer overflows, can easily undermine the security guarantees offered from internal.
Lefeuvre~\etal~\cite{lefeuvre:2023:towards} discuss the requirement for fast confidential IO, particularly focusing on the proper boundaries between the host and TEE I/O.

%% file: main.bbl
\begin{thebibliography}{10}

\bibitem{nginx:2017:nginx-perf}
{Testing the Performance of NGINX and NGINX Plus Web Servers}.
\newblock
  \url{https://www.nginx.com/blog/testing-the-performance-of-nginx-and-nginx-\plus-web-servers/},
  2017.

\bibitem{akram:2022:sok}
Ayaz Akram, Venkatesh Akella, Sean Peisert, and Jason Lowe-Power.
\newblock Sok: Limitations of confidential computing via tees for
  high-performance compute systems.
\newblock In {\em 2022 IEEE International Symposium on Secure and Private
  Execution Environment Design (SEED)}, pages 121--132. IEEE, 2022.

\bibitem{akram:2021:performance}
Ayaz Akram, Anna Giannakou, Venkatesh Akella, Jason Lowe-Power, and Sean
  Peisert.
\newblock Performance analysis of scientific computing workloads on general
  purpose tees.
\newblock In {\em 2021 IEEE International Parallel and Distributed Processing
  Symposium (IPDPS)}, pages 1066--1076. IEEE, 2021.

\bibitem{aws:2023:sev}
Amazon.
\newblock {Amazon EC2 now supports AMD SEV-SNP}.
\newblock
  \url{https://aws.amazon.com/about-aws/whats-new/2023/04/amazon-ec2-amd-sev-snp/},
  2023.

\bibitem{amd:2020:snp}
AMD.
\newblock {AMD SEV-SNP}: Strengthening {VM} isolation with integrity protection
  and more.
\newblock {\em White paper}, 2020.

\bibitem{amd:2020:ghcb}
AMD.
\newblock {SEV-ES} guest-hypervisor communication block standardization, 2020.

\bibitem{amd:2022:ciphertext}
AMD.
\newblock Technical guidance for mitigating effects of ciphertext visibility
  under amd sev.
\newblock
  \url{https://www.amd.com/system/files/documents/221404394-a_security_wp_final.pdf},
  2022.

\bibitem{amd:2023:tio}
{AMD}.
\newblock {AMD SEV-TIO: Trusted I/O for Secure Encrypted Virtualization}.
\newblock 2023.

\bibitem{amd:2023:github}
AMD.
\newblock {AMDSEV/sev-snp-devel} branch.
\newblock \url{https://github.com/AMDESE/AMDSEV/tree/sev-snp-devel}, 2023.

\bibitem{arm:2021:cca}
{ARM}.
\newblock {ARM CCA Security Model 1.0}, 2021.

\bibitem{arnautov:2016:scone}
Sergei Arnautov, Bohdan Trach, Franz Gregor, Thomas Knauth, Andre Martin,
  Christian Priebe, Joshua Lind, Divya Muthukumaran, Dan O'keeffe, Mark~L
  Stillwell, et~al.
\newblock {SCONE: Secure linux containers with intel SGX}.
\newblock In {\em 12th USENIX Symposium on Operating Systems Design and
  Implementation (OSDI 16)}, pages 689--703, 2016.

\bibitem{baidu:2023:dperf}
{Baidu}.
\newblock dperf.
\newblock \url{https://github.com/baidu/dperf}, 2023.

\bibitem{buhren:2017:fault}
Robert Buhren, Shay Gueron, Jan Nordholz, Jean-Pierre Seifert, and Julian
  Vetter.
\newblock Fault attacks on encrypted general purpose compute platforms.
\newblock In {\em 7th {ACM} on Conference on Data and Application Security and
  Privacy}. ACM, 2017.

\bibitem{buhren:2021:one}
Robert Buhren, Hans-Niklas Jacob, Thilo Krachenfels, and Jean-Pierre Seifert.
\newblock One glitch to rule them all: Fault injection attacks against amd's
  secure encrypted virtualization.
\newblock In {\em Proceedings of the 2021 ACM SIGSAC Conference on Computer and
  Communications Security}, pages 2875--2889, 2021.

\bibitem{confidential_consortium:2022}
{Confidential Computing Consortium}.
\newblock {Confidential Computing Consortium Members}.
\newblock \url{https://confidentialcomputing.io/members/}, 2022.

\bibitem{costan:2016:intel}
Victor Costan and Srinivas Devadas.
\newblock Intel {SGX} explained.
\newblock {\em IACR Cryptol. ePrint Arch.}, 2016(86):1--118, 2016.

\bibitem{dong:2012:sriov}
Yaozu Dong, Xiaowei Yang, Jianhui Li, Guangdeng Liao, Kun Tian, and Haibing
  Guan.
\newblock {High performance network virtualization with SR-IOV}.
\newblock {\em Journal of Parallel and Distributed Computing},
  72(11):1471--1480, 2012.

\bibitem{dpdk:2023:dpdk}
DPDK.
\newblock {Data Plane Development Kit}.
\newblock \url{https://www.dpdk.org/}, 2023.

\bibitem{du:2017:sevUnsecure}
Zhao-Hui Du, Zhiwei Ying, Zhenke Ma, Yufei Mai, Phoebe Wang, Jesse Liu, and
  Jesse Fang.
\newblock Secure encrypted virtualization is unsecure.
\newblock {\em arXiv preprint arXiv:1712.05090}, 2017.

\bibitem{fstack:2023:fstack}
{F-stack}.
\newblock F-stack.
\newblock \url{https://github.com/F-Stack/f-stack}, 2023.

\bibitem{google:2020:sev}
Google.
\newblock {Introducing Google Cloud Confidential Computing with Confidential
  VMs}.
\newblock
  \url{https://cloud.google.com/blog/products/identity-security/introducing-google-cloud-confidential-\\computing-with-confidential-vms},
  2020.

\bibitem{hetzelt:2017:security}
Felicitas Hetzelt and Robert Buhren.
\newblock Security analysis of encrypted virtual machines.
\newblock In {\em ACM SIGPLAN Notices}. ACM, 2017.

\bibitem{hetzelt:2021:via}
Felicitas Hetzelt, Martin Radev, Robert Buhren, Mathias Morbitzer, and
  Jean-Pierre Seifert.
\newblock Via: Analyzing device interfaces of protected virtual machines.
\newblock In {\em Annual Computer Security Applications Conference}, pages
  273--284, 2021.

\bibitem{intel:2020:tdx}
Intel.
\newblock Intel trust domain extensions whitepaper.
\newblock
  \url{https://software.intel.com/content/dam/develop/external/us/en/documents/tdx-whitepaper-final9-17.pdf},
  2020.

\bibitem{intel:2023:qat}
Intel.
\newblock {Intel QuickAssist Technology (Intel QAT)}.
\newblock 2023.

\bibitem{intel:2023:tio}
Intel.
\newblock {Intel TDX Connect TEE-IO Device Guide}.
\newblock 2023.

\bibitem{kaffes:2019:shinjuku}
Kostis Kaffes, Timothy Chong, Jack~Tigar Humphries, Adam Belay, David
  Mazi{\`e}res, and Christos Kozyrakis.
\newblock {Shinjuku: Preemptive scheduling for $\mu$second-scale tail latency}.
\newblock In {\em 16th USENIX Symposium on Networked Systems Design and
  Implementation (NSDI 19)}, pages 345--360, 2019.

\bibitem{kaplan:2017:seves}
David Kaplan.
\newblock Protecting {VM} register state with {SEV-ES}.
\newblock {\em White paper}, 2017.

\bibitem{kaplan:2016:sevWpaper}
David Kaplan, Jeremy Powell, and Tom Woller.
\newblock {AMD} memory encryption.
\newblock {\em White paper}, 2016.

\bibitem{lefeuvre:2023:towards}
Hugo Lefeuvre, David Chisnall, Marios Kogias, and Pierre Olivier.
\newblock {Towards (Really) Safe and Fast Confidential I/O}.
\newblock In {\em Proceedings of the 19th Workshop on Hot Topics in Operating
  Systems}, pages 214--222, 2023.

\bibitem{li:2023:analysis}
Dingji Li, Zeyu Mi, Chenhui Ji, Yifan Tan, Binyu Zang, Haibing Guan, and Haibo
  Chen.
\newblock Analysis and optimization of network i/o tax in confidential virtual
  machines.
\newblock In {\em Proceedings of the 2023 USENIX Conference on Usenix Annual
  Technical Conference}.

\bibitem{li:2022:systematic}
Mengyuan Li, Luca Wilke, Jan Wichelmann, Thomas Eisenbarth, Radu Teodorescu,
  and Yinqian Zhang.
\newblock {A Systematic Look at Ciphertext Side Channels on AMD SEV-SNP}.
\newblock In {\em 2022 IEEE Symposium on Security and Privacy (SP)}, pages
  1541--1541. IEEE Computer Society, 2022.

\bibitem{li:2020:crossline}
Mengyuan Li, Yinqian Zhang, and Zhiqiang Lin.
\newblock {CROSSLINE: Breaking ``Security-by-Crash'' based Memory Isolation in
  AMD SEV}.
\newblock {\em arXiv preprint arXiv:2008.00146}, 2020.

\bibitem{li:2021:cipherleaks}
Mengyuan Li, Yinqian Zhang, Huibo Wang, Kang Li, and Yueqiang Cheng.
\newblock {CIPHERLEAKS: Breaking Constant-time Cryptography on AMD SEV via the
  Ciphertext Side Channel}.
\newblock In {\em 30th USENIX Security Symposium (USENIX Security 21)}, pages
  717--732, 2021.

\bibitem{li:2021:tlb}
Mengyuan Li, Yinqian Zhang, Huibo Wang, Kang Li, and Yueqiang Cheng.
\newblock Tlb poisoning attacks on amd secure encrypted virtualization.
\newblock In {\em Annual Computer Security Applications Conference}, pages
  609--619, 2021.

\bibitem{microsoft:2021:sev}
Microsoft.
\newblock {Azure and AMD announce landmark in confidential computing
  evolution}.
\newblock
  \url{https://azure.microsoft.com/en-us/blog/azure-and-amd-enable-lift-and-shift-\\confidential-computing/},
  2021.

\bibitem{morbitzer:2019:extract_new}
Mathias Morbitzer, Manuel Huber, and Julian Horsch.
\newblock Extracting secrets from encrypted virtual machines.
\newblock In {\em 9th {ACM} Conference on Data and Application Security and
  Privacy}. ACM, 2019.

\bibitem{morbitzer:2018:severed}
Mathias Morbitzer, Manuel Huber, Julian Horsch, and Sascha Wessel.
\newblock {SEVered}: Subverting {AMD}'s virtual machine encryption.
\newblock In {\em 11th European Workshop on Systems Security}. ACM, 2018.

\bibitem{morbitzer:2021:severity_new}
Mathias Morbitzer, Sergej Proskurin, Martin Radev, Marko Dorfhuber, and
  Erick~Quintanar Salas.
\newblock Severity: Code injection attacks against encrypted virtual machines.
\newblock In {\em 2021 IEEE Security and Privacy Workshops (SPW)}, pages
  444--455. IEEE, 2021.

\bibitem{nginx:2023:nginx}
Nginx.
\newblock {NGINX: Advanced Load Balancer, Web Server, \& Reverse Proxy}.
\newblock \url{https://www.nginx.com/}, 2023.

\bibitem{orenbach:2017:eleos}
Meni Orenbach, Pavel Lifshits, Marina Minkin, and Mark Silberstein.
\newblock {Eleos: ExitLess OS services for SGX enclaves}.
\newblock In {\em Proceedings of the Twelfth European Conference on Computer
  Systems}, pages 238--253, 2017.

\bibitem{priebe:2019:sgx-lkl}
Christian Priebe, Divya Muthukumaran, Joshua Lind, Huanzhou Zhu, Shujie Cui,
  Vasily~A Sartakov, and Peter Pietzuch.
\newblock {SGX-LKL: Securing the host OS interface for trusted execution}.
\newblock {\em arXiv preprint arXiv:1908.11143}, 2019.

\bibitem{russell:2008:virtio}
Rusty Russell.
\newblock {virtio: towards a de-facto standard for virtual I/O devices}.
\newblock {\em ACM SIGOPS Operating Systems Review}, 42(5):95--103, 2008.

\bibitem{thalheim:2021:rktio}
J{\"o}rg Thalheim, Harshavardhan Unnibhavi, Christian Priebe, Pramod Bhatotia,
  and Peter Pietzuch.
\newblock {Rkt-io: A direct i/o stack for shielded execution}.
\newblock In {\em Proceedings of the Sixteenth European Conference on Computer
  Systems}, pages 490--506, 2021.

\bibitem{van:2017:sgx}
Jo~Van~Bulck, Frank Piessens, and Raoul Strackx.
\newblock {SGX-S}tep: A practical attack framework for precise enclave
  execution control.
\newblock In {\em Proceedings of the 2nd Workshop on System Software for
  Trusted Execution}, pages 1--6, 2017.

\bibitem{wang:2023:pwrleak}
Wubing Wang, Mengyuan Li, Yinqian Zhang, and Zhiqiang Lin.
\newblock Pwrleak: Exploiting power reporting interface for side-channel
  attacks on amd sev.
\newblock In {\em International Conference on Detection of Intrusions and
  Malware, and Vulnerability Assessment}, pages 46--66. Springer, 2023.

\bibitem{werner:2019:severest}
Jan Werner, Joshua Mason, Manos Antonakakis, Michalis Polychronakis, and Fabian
  Monrose.
\newblock The {SEV}er{ESt} of them all: Inference attacks against secure
  virtual enclaves.
\newblock In {\em {ACM} Asia Conference on Computer and Communications
  Security}, pages 73--85. ACM, 2019.

\bibitem{wrk:2023:github}
WG.
\newblock wrk - a http benchmarking tool.
\newblock \url{https://github.com/wg/wrk/}, 2023.

\bibitem{wilke:2021:undeserved}
L.~Wilke, J.~Wichelmann, F.~Sieck, and T.~Eisenbarth.
\newblock undeserved trust: Exploiting permutation-agnostic remote attestation.
\newblock In {\em 2021 IEEE Security and Privacy Workshops (SPW)}, pages
  456--466, Los Alamitos, CA, USA, may 2021. IEEE Computer Society.

\bibitem{wilke:2020:sevurity}
Luca Wilke, Jan Wichelmann, Mathias Morbitzer, and Thomas Eisenbarth.
\newblock {SEV}urity: No security without integrity: Breaking integrity-free
  memory encryption with minimal assumptions.
\newblock In {\em 2020 IEEE Symposium on Security and Privacy (SP)}, pages
  1483--1496. IEEE, 2020.

\end{thebibliography}
